\documentclass[a4paper,11pt]{article}

\usepackage{setspace}
\usepackage[latin1]{inputenc}
\usepackage[english]{babel}
\usepackage{graphicx, subfigure}
\usepackage{amsmath, amsfonts, amssymb, latexsym}
\usepackage{float,latexsym, basicsymbols, notation}
\usepackage{cite}
\usepackage{color}
\usepackage{authblk}
\usepackage[pdfborder={0 0 0}]{hyperref}

\author[1]{Adri\`a Tauste Campo}
\author[1,2,3]{Marina Martinez-Garcia}
\author[4]{Ver\'onica N\'acher}
\author[4]{Rogelio Luna}
\author[4,5]{Ranulfo Romo}
\author[1,6]{Gustavo Deco}
\vspace{-3mm}
\affil[1]{\small{Center for Brain and Cognition, Department of Information and Communication Technologies, Universitat Pompeu Fabra, 08018 Barcelona, Spain} }
\affil[2]{\small{Department of Ophthalmology, RWTH Aachen University, 52076 Aachen, Germany}}
\affil[3]{\small{Institute of Neuropathology, RWTH Aachen University, 52076 Aachen, Germany}}
\affil[4]{\small{Instituto de Fisiología Celular-Neurociencias, Universidad Nacional Aut\'onoma de M\'exico, 04510 Mexico D.F., Mexico}}
\affil[5]{\small{El Colegio Nacional, 06020 Mexico D.F., Mexico}}
\affil[6]{\small{Instituci\'o Catalana de Recerca i Estudis Avan{\c c}ats, Passeig Lluís Companys, 23, 08010 Barcelona, Spain}}

\title{Task-driven intra- and interarea communications in primate cerebral cortex}

\begin{document}
\date{}
\maketitle

\begin{abstract}
Neural correlations during a cognitive task are central to study brain information processing and computation. However, they have been poorly analyzed due to the difficulty of recording simultaneous single neurons during task performance. In the present work, we quantified neural directional correlations using spike trains that were simultaneously recorded in sensory, premotor, and motor cortical areas of two monkeys during a somatosensory discrimination task. Upon modeling spike trains as binary time series, we used a nonparametric Bayesian method to estimate pairwise directional correlations between many pairs of neurons throughout different stages of the task, namely, perception, working memory, decision making, and motor report. We find that solving the task involves feedforward and feedback correlation paths linking sensory and motor areas during certain task intervals. Specifically, information is communicated by task-driven neural correlations that are significantly delayed across secondary somatosensory cortex, premotor, and motor areas when decision making takes place. Crucially, when sensory comparison is no longer requested for task performance, a major proportion of directional correlations consistently vanish across all cortical areas.
\end{abstract}

%
\section{Introduction}

The problem of neural communication in the brain 
has been traditionally little explored 
due to the need for simultaneous recordings \cite{Salinas01}.
The 
arrival of new techniques  to record both
neural population activity and single-neuron
action potentials
offers 
new prospects to study this problem  \cite{Brown04,Buz04}. 
Recently, population recordings have motivated  a large number of works on multi-unit interactions, 
including the study of interactions between local field potentials (LFP) \cite{Roelf97,Wom07,Nac13}; LFP and multi-unit activity (MUA) \cite{Wom07};
and LFP and neuronal spikes \cite{Kor13} but
lesser attention has been payed at interactions between single-unit recordings \cite{Hoff02}.  
However, the analysis of simultaneous spike trains  becomes critical 
as it is generally assumed that
neurons are key units in distributing information
across  brain areas \cite{Kolb01}.

An ideal paradigm to study neural communication is the so-matosensory discrimination task designed by Romo and coworkers 
\cite{Her97}.  
 In this task, a trained monkey discriminates the difference 
in frequency between two mechanical vibrations delivered sequentially to one fingertip (Fig.~1A). 
Essentially, the monkey must hold the first stimulus frequency ($f1$) in working memory, must compare 
the second stimulus frequency ($f2$) to the memory trace of $f1$ to form a decision of whether $f2 > f1$ or $f2 < f1$, 
and must postpone the decision until a sensory cue triggers the motor report \cite{Lem07}.
At the end of every trial the monkey is rewarded with a drop of liquid for correct discriminations.
Previous work on this task has analyzed how single-neuron responses across sensory and motor areas linearly 
correlate with stimuli and the decision report during the key stages of the task  \cite{Her10}. 
The results show that stimuli are mostly encoded in somatosensory areas, the processes of working memory; 
that comparison takes place in the secondary somatosensory cortex (S2) and premotor areas; and that behavioral information is primarily found in premotor and motor areas. Thus, the somatosensory discrimination task activates complex processes that are required to communicate information from the areas that encode the stimuli to the areas that integrate them and report the decision.

In the present work, we study this communication paradigm through the analysis of simultaneous recordings of neurons engaged in the task \cite{Nac13,Her08} from two monkeys. Indeed, by applying nonlinear statistical methods, we estimate modulated cortical correlations that help describe how task-related information flows from sensory to motor areas when a correct decision is made.

%

\section{Results}

We studied interactions between neuronal spike trains that were simultaneously recorded from five cortical areas in one trained monkey performing a somatosensory discrimination task \cite{Nac13}. Recordings for the first monkey were performed in $13$ independent sessions ($n = 13$). During each session, up to seven microelectrodes were individually inserted in each of the five cortical areas for simultaneous recordings of single neurons. The selected neurons were from two somatosensory areas, primary somatosensory cortex (S1) and S2, and three premotor/motor areas, medial premotor cortex (MPC), dorsal premotor cortex (DPC), and primary motor cortex (M1) (Materials and Methods). To investigate neural correlations during the discrimination task, we only considered correct (``hit") trials of similar psychophysical performance. We validated our results by repeating the same analysis on a second trained monkey in $19$ sessions divided into two block of three 
simultaneously recorded areas each (S1, S2, and DPC and S1, S2, and M1).

The central  measure of our analysis is the directed information (DI), mathematically denoted by $I(X^T\rightarrow Y^T)$, which is a non-linear measure of 
directional correlation
 between 
 the processes $X^T$ and $Y^T$ \cite{Mass90}. When the two process are identical (i.e.,
  $X^T=Y^T$), this measure coincides with the 
  Shannon entropy, denoted by $H(Y^T)$  \cite{Shannon49}.
   The DI $I(X^T\rightarrow Y^T)$ quantifies for any given time $t$ the information that the past and present of $X^T$ (up to time $t$) 
   has about the present of $Y^T$ upon the knowledge of the past of $Y^T$ (up to time $t-1$). 
 Alternatively, the entropy of $Y^T$ quantifies the uncertainty on any realization of $Y^T$. 

\subsection{Neural correlations are task driven}

We estimated the entropy in $870$ neurons and the DI in $50616$ ordered neuron pairs from two monkeys to infer significant auto- and pairwise directional correlations across time delays of $0,10, 20, \cdots, 140$ ms and during $17$ consecutive task intervals of $0.5$ s, spanning from the interval before the $f1$ stimulation to the interval after the lift of the sensory cue. This cue lift interval will hereafter be referred to as the probe-up (pu) period (Fig. 1A and Materials and Methods).
In particular, for every task interval, we tested the significance of each estimated measure against a null hypothesis of complete directional independence using the maximum value of the DI over all selected delays as a test statistic (Materials and Methods). First, we selected every neuron whose entropy was significant (permutation test, $\alpha = 5\%$) for at least one of the frequency pairs and denoted it a ``responsive neuron".  In a similar vein, every significant correlation between responsive neurons for at least one of the frequency pairs was denoted a ``responsive path" and each correlated neuron was denoted either a ``starting point" or ``end point" 
neuron according to the correlation's directionality. Conversely, every path whose starting point/end point was a given neuron was denoted an ``outgoing/incoming" 
path to that neuron, respectively (SI Appendix, section 2). For every interarea comparison, we computed the percentage of responsive paths over all possible simultaneous pairs. For instance, SI Appendix, Fig. S1 shows that responsive paths in the first monkey were found above significance level ($\alpha' = 9.75\%$) across all area pairs and task intervals (green curves).

We studied whether responsive paths were directly associated with the discrimination task. To this end, we estimated again the directional correlation in every neuron pair that formed a responsive path during a control task, in which the monkey received identical mechanical vibrations but was requested to remain still upon a reward that arrived at variable time (passive stimulation; Fig. 1A and Materials and Methods). Under passive stimulation, only a small fraction of the responsive paths were again found significant ($\approx20\%$) across all area pairs and task intervals (gray curves, SI Appendix, Fig. S1). Overall, there was a low correlation between the presence of responsive paths during the original task and the control task in both monkeys ($\rho<0.03$, Spearman correlation \cite{Ken48}), suggesting that neural correlations were driven by weakly dependent processes.

We then wondered whether both tasks were also differentiated by single-neuron measures under fixed stimulation. To examine this question, we focused on the ensemble of neurons that were end point neurons of responsive paths and measured their activity during each task. First, we measured the firing rate and spike-train entropy of every neuron in the ensemble. Then, as a benchmark multineuron measure, we evaluated the aggregated sum of DI along every neuron's incoming responsive paths. For the first monkey, Fig. 2 shows the average firing rate, average entropy, and average incoming DI during both tasks when $(f1=14,f2=22)$ Hz, 
together with their error bars. Neurons exhibited consistent single-neuron differences across tasks in MPC, DPC, and M1 around $f2$ stimulation.

In contrast, the use of directional correlations shows that in those periods when neurons were equally firing in both tasks (or with similar spike-train entropies), they were less influenced through responsive paths during passive stimulation than during somatosensory discrimination. Results were similar for the frequency pair 
$(f1 = 30, f2 = 22)$ Hz (SI Appendix, Fig. S2) and were also corroborated in a second monkey (SI Appendix, Fig. S7).

 \subsection{Neural correlations are modulated by decision making}

 Despite being task-driven, the role of responsive paths still remained unclear because they were uniformly present across all areas and task intervals (SI Appendix, Fig. S1). In particular, to what extent were these paths communicating task-related information?
 
To investigate more intrinsic connections between neural correlations and decision making, we searched for the subset of responsive paths that were significantly modulated by a key task variable. More precisely, we tested the modulation of every responsive neuron and path with respect to the decision sign $D=f1-f2$ by computing the difference between the DI estimates across trials recorded at frequency pairs $(f1=14, f2=22)$ Hz ($D<0$) and $(f1=30, f2=22)$ Hz ($D>0$). Responsive neurons and paths that were significantly modulated (permutation test, $\alpha = 5\%$) were thus denoted as ``modulated neurons" and ``modulated paths" respectively. By this choice of trials, modulated paths have a different interpretation depending on the task interval. For instance, during the intervals before $f2$ stimulation, modulations can be regarded as correlates of $f1$, whereas during the intervals after $f2$ stimulation, they can be interpreted as correlates of the decision sign and the associated motor action. Green curves in Fig. 3 and SI Appendix, Fig. S3 show the percentage of modulated neurons (Fig. 3A) and modulated paths (Fig. 3B and SI Appendix, Fig. S3) while the first monkey was performing the discrimination task, and black circles indicate the intervals when this percentage was significantly (Agresti- Coull confidence interval (17),  \cite{Ag98}, $\alpha = 5$) above chance level ($\alpha= 5\%$). In the rest of this article, we will use this statistical sense when referring to percentages that are above significant level. Area comparisons in Fig. 3 were chosen to describe the chain of comparisons S1$\leftrightarrow$ S2$\leftrightarrow$ 
 MPC $\leftrightarrow$ DPC $\leftrightarrow$M1. In general, Fig. 3 and SI Appendix, Fig. S3 show that the directional measure was able to discriminate top-down from bottom-up interactions in each recorded area pair.

Because both modulated neurons and paths carried task in- formation, we analyzed their mutual relationship by computing the proportion of modulated paths that linked modulated neurons. In contrast to responsive paths, modulated neurons were positively correlated with the presence of their own outgoing or incoming modulated path in both monkeys (Spearman correlation; SI Appendix, Figs. S4A and S9A), which implied that the proportion of modulated paths linking modulated neurons was above chance level for every area and during the majority of task intervals (first monkey; SI Appendix, Fig. S4B). This preliminary result indicates that modulated paths were prone to link modulated neurons.
  
 Fig. 3 and SI Appendix, Fig. S3 illustrate how sensory information was encoded and distributed from sensory to motor areas in the first monkey while the percept was processed to drive a motor action. On one hand, S1 encoded $f1$ at the first stimulation period and was especially active in distributing this information toward S2, MPC, and M1 during working memory intervals (Fig. 3 and SI Appendix, Fig. S3). On the other hand, S1 was interactive with premotor and motor areas around the pu period, which suggests that the sensory cue delivered at the pu period produced sensory-motor correlations that could anticipate the report (SI Appendix, Fig. S3). Neurons from S2 received sensory information from S1 and MPC during the interstimulus interval (Fig. 3B) but did not individually encode it (Fig. 3A). This lack of local activity after $f1$ stimulation may be a consequence of S2's function in encoding and integrating $f2$ (12), whose value was kept fixed in our study. The role of MPC was to mediate between sensory and motor areas in two different stages. First, during the intervals before $f2$ stimulation, MPC received incoming interactions from sensory areas that were modulated by $f1$ (Fig. 3B and SI Appendix, Fig. S3). Second, during the post- poned decision, MPC mainly produced outgoing interactions to M1 and to other neurons within MPC that were modulated by the decision sign (SI Appendix, Fig. S3). Neurons from DPC showed heterogeneous communication patterns with respect to sensory areas and MPC. However, a great proportion of them received task information by persistent links from M1. The area M1 exhibited two distinct roles while communicating with the rest of cortical areas. First, it was influenced by S1 and S2 around $f2$ stimulation, which indicates that there was an information link between sensory and motor areas regardless of the encoding patterns found in each area (Fig. 3A and SI Appendix, Fig. S3). Second, M1 was highly interactive with the rest of premotor and motor areas in the process of conforming the decision and the motor action in line with previous results \cite{Her10}. The pattern of intraarea activity of M1 peaked at $f2$ stimulation and near the pu period, which indicates that these events were the main drivers of local information exchange.
 
In contrast, during passive stimulation, neither modulated neurons nor modulated paths were generally associated with $f1$ and the monkey's choice. Modulated neurons and paths only showed some persistency in S1 during the first stimulation (gray curves, Fig. 3), which confirms the existence of minimal sensory processing in S1 during the control task \cite{Her10} . This abrupt change in the activity of modulated neurons and paths was corroborated in a second monkey (SI Appendix, Fig. S8).

 \subsection{Task-specific delayed correlations distribute sensory and behavioral information}

 We further studied interarea communications by analyzing two characteristics of modulated paths: their modulation rule and their correlation delay. First, we divided modulated paths into three classes: ON-ON, defined to involve significant correlations in both decision signs, and ON-OFF and OFF-ON, defined to be significant only for $f1< f2$ and $f1 > f2$, respectively. The majority of modulated paths in both monkeys were of the form ON-OFF and OFF-ON (Fig. 4A and SI Appendix, Fig. S9B), indicating that task information was mainly encoded by the presence of significant correlations during trials of one decision report that vanished during trials of the opposite decision report. The almost equal contribution of ON-OFF and OFF-ON modulations along the task gave rise to an overall picture that was difficult to interpret (SI Appendix, Fig. S5). To observe how these modulations interplayed for a specific interarea comparison in the first monkey, Fig. 4B shows the percentage of 
 modulation classes above significant level ($\alpha = 5\%$) and Fig. 4C shows the sum of the average (across trials) DI along modulated paths starting at M1 and ending at DPC in each decision report. Fig. 4 B and C also highlight the three stages where the aggregated DI peaked (time periods and corresponding values in dashed rect- angles). These stages may be linked to the acquisition of $f1$ (intervals 2 and 3), the recovery of $f1$ before the comparison takes place (interval 9), and the process of planning the action (interval 15). Comparison of both figures at the later stage shows that a similar number of sign-specific paths could lead to twice as much aggregated DI for one decision report than for the opposite decision report.

To study interneuronal delays, we divided the modulated paths that were found above significant level in the first monkey into three sets according to their estimated delays: $0$ ms, $[10,70]$ ms, and $[80,140]$ ms. The interneuronal interactions at these delays were computed over a correlation memory of $4$ ms (Materials and Methods), thus capturing effects additional to those effects found using classical synchronization measures \cite{Vic05}. We first plotted the distribution of delays across area pairs and task intervals where modulated paths were above significant level (SI Appendix, Fig. S6; $\alpha=5\%$). We summarized these findings in Fig. 5 after having classified interneuronal delays according to the function (somatosensory/premotor/motor) and location (right/left hemisphere) of each area under comparison.

 Overall, modulated paths across somatosensory areas were dominated by instantaneous interactions, whereas modulated paths involving premotor and motor areas were mostly delayed at the range of $[80,140]$ ms (Fig. 5A). In particular, we tested the average delay across area pairs and obtained significant differences 
 (Fig. 5B; Wilcoxon test \cite{Ken48},  $p < 0.005$) between somatosensory interactions ($43.6$ ms); interactions between S1 and MPC, DPC, and M1 ($64.2$ ms); and interactions across S2, MPC, DPC, and M1 ($73.4$ ms). These differences also emerged after removing the contribution of instantaneous correlations but were no longer significant (Fig. 5C; Wilcoxon test \cite{Ken48}, $p>0.05$). Further, a closer look at SI Appendix, Fig. S6 reveals that modulated paths within the somatosensory cortex and within M1 were less delayed than interactions across premotor and motor areas. These findings suggested that differences in the average delay could be driven by the relative location of the areas in the two hemispheres. Then, we computed the average delay across areas within a hemisphere and across areas from distinct hemispheres, obtaining significant differences that were robust to the effect of intraarea correlations (Fig. 5B; Wilcoxon test \cite{Ken48}, $p < 0.005$). However, this delay difference did not remain significant after removing the effect of instantaneous correlations (Fig. 5C; Wilcoxon test \cite{Ken48}, $p > 0.05$). In sum, instantaneous correlations were key to discriminate interarea relationships with respect to the areas'  location and function.

\section{Discussion}

Using nonparametric estimation of spike-train interdependencies, we have unraveled neural correlation paths that are specific to a discrimination task. These paths are task-driven for two main reasons. First, they dramatically decrease when the monkey receives both stimuli but is not requested to perform the cognitive task (Fig. 2). Second, they are modulated in a significant percentage by sensory and behavioral variables (Fig. 3). More importantly, these modulated paths are related to neurons that individually encode task variables and are therefore likely to distribute their information further across other areas (SI Appendix, Fig. S4). In general, the use of directional correlations seems to discriminate the original task from a control task better than single-neuron measures (Fig. 2A), suggesting that task-driven correlations may not generally be rate-dependent \cite{Rocha07}.

Modulated paths can be used to characterize the role of each area in distributing relevant information to solve the task  \cite{Lem07, Her10}. In particular, we observed that S1 is particularly important in feed-forwarding sensory information to superior areas, S2 interacts with MPC during the working memory stage, MPC acts as a relay node between sensory and motor areas, and the interactions across MPC $\leftrightarrow$  M1 $\rightarrow$ DPC concentrate the information on the monkey's choice (Fig. 3 and SI Appendix, Fig. S3). Modulated paths mainly encode task information by the existence (ON) and absence (OFF) of a given neural correlation, which in particular indicates that each decision is distributed across areas through a different subset of interactions (Fig. 4). For each decision report, modulation paths are delayed according to the hemisphere location and function of each area under comparison. In particular, modulated paths are significantly faster when they distribute 
information across somatosensory areas (S1 and S2) during intervals before the $f2$ stimulation than when they link S2, premotor, and motor areas during decision making. These findings indicate that sensory and behavioral information may be communicated at different time scales.

Our description of modulated paths in primate cerebral cortex extends beyond previous works in which task-related activity was found to be highly distributed across areas and time intervals  \cite{Her10}. To encompass both sets of results, we make hypotheses in two related directions. On the one hand, our analysis of modulation paths suggests that task-related information is jointly encoded by neurons that often do not exhibit individual modulations. On the other hand, the great percentage of responsive paths that are not modulated indicates that there is context- dependent activity beyond encoding of sensory and behavioral variables ($f1$, $f2$, and decision). This activity may be the result of internal processes involving sensory and motor areas, such as arousal, attention, or motivation \cite{Cohen11}, whose encoding patterns could not be captured with this experimental paradigm. The study of these hypotheses may shed light on the underlying mechanisms that encode, distribute, and transform the required information to solve a cognitive task.

\section{Materials and Methods}
This study was performed on two adult male monkeys (\emph{Macaca mulatta}), weigthing
8-12 kg. All procedures followed the guidelines of the National Institutes of Health
and Society for Neuroscience. Protocols were approved by the Institutional 
Animal Care and Use Committee of the Instituto de Fisiolog\'ia Celular.

\subsection{Recordings}
Data acquisition, amplification, and filtering were described in
detail \cite{Her08}. In brief, the activity of single neurons were simultaneously
recorded with an array of seven independent, movable microelectrodes
(1-1.5 M$\Omega$) inserted in each of five cortical areas. Electrodes
within an area were spaced 305 or 500 $\mu$m apart \cite{Eck93}. Spike sorting was
performed manually on-line, and single neurons were selected if they
responded to any of the different components of the discrimination task.
In particular, neurons from area S1 had cutaneous receptive fields with quickly adapting (QA) properties, 
whereas those from area S2 had large cutaneous receptive fields with no obvious submodality properties. Neurons of the 
frontal cortex had no obvious cutaneous or deep receptive fields; they were selected if they responded to any of the different 
components of the discrimination task
\cite{Her10,Her08}. The cortical areas were the S1, S2, MPC, DPC, and M1 (Fig. 1B).
Recordings in S1, S2, and DPC were made in the hemisphere contralateral to
the stimulated hand (left hemisphere), and in MPC and M1 contralateral to the
responding hand/arm (right hemisphere). 

\subsection{Discrimination Task} The paradigm used here has been described \cite{Her97, Lem07}. The
monkey sat on a primate chair with its head fixed in an isolated, sound-
proof room. The right hand was restricted through a half-cast and kept in
a palm-up position. The left hand operated an immovable key (elbow at
$\sim 90^\circ$ ), and two push buttons were in front of the animal, $25$cm away from
the shoulder and at eye level. The centers of the switches were located $7$ and
$10.5$cm to the left of the midsagittal plane. In all trials, the monkey first
placed the left hand and later projected to one of the two switches. Stimuli
were delivered to the skin of the distal segment of one digit of the right,
restrained hand, via a computer-controlled stimulator ($2$mm round tip; BME
Systems). The initial probe indentation was $500\mu$m. Vibrotactile stimuli were
trains of short mechanical pulses. Each of these pulses consisted of a single-
cycle sinusoid lasting $20$ms. Stimulus amplitudes were adjusted to equal
subjective intensities; for example, $71\mu$m at $12$Hz and $51\mu$m at $34$Hz (a
decrease of $\sim 1.4\%$ per Hz). During discrimination trials (Fig. 1A), the 
mechanical probe was lowered (probe down; pd), indenting the glabrous skin
of one digit of the hand; the monkey placed its free hand on an immovable
key (key down; kd); after a variable prestimulus delay ($0.5-3$s) the probe
oscillated vertically at the frequency of the first stimulus ($f1$); after a fixed
delay ($3$s), a second mechanical vibration was delivered at the second
stimulus ($f2$) frequency; after another fixed delay ($3$s) the probe was lifted off
from the skin (probe up; \emph{pu}); the monkey released the key (\emph{ku}) and pressed
either a lateral or a medial push button (\emph{pb}) to indicate whether $f2$ was of
higher or lower frequency than $f1$, respectively. The monkey was rewarded
with a drop of liquid for correct discriminations. 
The experimental sets of frequency pairs used during the discrimination task were the same as in \cite{Her10}, for both monkeys.

\subsection{Control tests}
 During a passive stimulation condition, the monkey was trained to maintain its free arm motionless during the trial (Fig. 1B). Stimuli were delivered to the fingertip, and the animal remained alert by being rewarded with drops of liquid at different times, but no motor response with the free hand was required.

\subsection{Data analysis}
Data were analyzed offline by using costume-build MATLAB code (MathWorks).
We selected $13$ experimental sessions from the first monkey and $19$ experimental sessions of the second monkey 
according to the following criteria. First, 
we selected 
sessions in which the monkey had similar psychophysical thresholds \cite{Her97}.
Second, our analysis required the existence of passive stimulation sessions registered on the population. 
We estimated neural directional correlations between every neuron pair 
within a population using a non-parametric estimator of  the DI between
a pair of discrete time series  that were assumed to be generated according to a Markovian process \cite{Jiao13}. In more specific terms,  
for a pair time series $(x_1^T,y_1^T)$ of length $T$, where $x_1^T=(x_1,\dots, x_T)$ and  $y_1^T=(y_1,\dots, y_T)$, a time delay $\delta\geq 0$  
and Markovian orders equal to $D_1>0$ and $D_2>0$ respectively,  
the DI
 between  the stationary processes of $x_1^T$ and $y_1^T$, i.e.,
$\mathcal{X}$ and $\mathcal{Y}$, 
%
%
is estimated  
through the formula
 \begin{align}
\hat{I}_{\delta}(\mathcal{X} \rightarrow \mathcal{Y})  
\triangleq 
 &\frac{1}{T}\sum_{t=1}^T \sum_{y_t} \hat{P}(Y_t=y_t\bigl |X^{t-\delta}_{t-\delta-D_2}=x^{t-\delta}_{t-\delta-D_2}, Y^{t-1}_{t-D_1}=y^{t-1}_{t-D_1})\notag\\
 & \times \log \frac{\hat{P}(Y_t=y_t \bigl| X^{t-\delta}_{t-\delta-D_2}=x^{t-\delta}_{t-\delta-D_2}, Y^{t-1}_{t-D_1}=y^{t-1}_{t-D_1})}
 {\hat{P}(Y_t=y_t\bigl|Y^{t-1}_{t-D_1}=y^{t-1}_{t-D_1})},\label{eqform2}
  \end{align}
where the probability distribution of $(X_1^T, Y_1^T)$ is estimated using the context-tree weighting algorithm (CTW) \cite{Will95}. 
Eq.  
\eqref{eqform2} quantifies the information that the past of $X_1^T$
at delay $\delta$ , i.e., $X^{t-\delta}_{t-\delta-D_2}$, 
has about the present of $Y_1^T$, i.e., $Y_t$, 
given the most recent past of $Y_1^T$
, i.e., $Y^{t-1}_{t-D_1}$.
This estimator is consistent 
as long as $(\mathcal{X},\mathcal{Y})$  
the two neuronal time series form
a jointly stationary irreducible aperiodic finite-alphabet
Markov process whose order does not exceed the
prescribed maximum depth in the CTW algorithm \cite[Th. 3]{Jiao13}.
Prior to 
estimate the DI
we preprocessed our data 
as follows.
For a fixed stimulation pair, we first binarized spike-train trials using bins of $2$ms (mapping $1$ 
to each bin with at least one spike and $0$, otherwise). We then divided each time series  into $17$ consecutive task intervals
of $0.5$s ($250$ bins). For each neuron, segments which corresponded to the same type of trials and task interval were assumed to be 
generated by a common  random process that satisfied the estimator requirements with a
maximum memory of $4$ms ($D_1=D_2=2$ bins) both for the joint and the marginal spike-train processes. 
Under these assumption, it can be easily checked that the DI is asymptotically equivalent to the transfer entropy 
\cite{Sch00}
in the limit of the time-series length.   
To assess that neurons were able to express
information through their spike-train responses, we run the estimator of the entropy (a particular case of the 
DI estimator) for each neuron and task interval
over the time series that resulted from the concatenation of the fixed stimulation trial  segments.  
Finally, among those neurons that had a significant entropy value, 
we run the DI 
estimator (See SI Appendix, Sec. 3) over all possible simultaneous neuron pairs 
across  delays $0,10,20, 30,40,50,60,70,80,90,100,110,120,130,140$ms, where the total range 
 was chosen to be compatible  
 with the latency 
 of each area \cite{Lem07,Laf06} (See SI Appendix, Sec. 4).

To assess the statistical significance of the estimations
 we used a Monte-Carlo permutation test \cite{Ernst04}, where
 the original (i.e., non permuted) results were compared with the tail of a distribution obtained by permuting $20$ times 
 the concatenations of the second binarized spike train $Y^T$ differently  for each original estimation ($\alpha=5\%$) and computed the corresponding p-value \cite{Phipson10}.  
  We dealt with the multiple test problem (one test for each delay) by using 
  the maximum DI over all selected delays as a test statistic. 
Further details about the significance analysis 
for the DI computations  and the
 modulation tests 
 are provided in the SI Appendix, Sec. 5.

\section{Acknowledgments}
R.R.'s research was partially supported by
International Research Scholars Award 55005959 from the Howard Hughes
Medical Institute, Direcci\'on de Personal Acad\' emico de la Universidad Nacional
Aut\'onoma de M\'exico Grant IN203210 and Consejo Nacional de Ciencia y
Tecnolog\'ia Grant CB-2009-01-130863. V.N. was supported by a Ministerio de
Educaci\'on y Ciencia (MEC)-Fullbright Postdoctoral Fellowship from the Spanish
Ministry of Science and Technology and Direcci\'on de Personal Acad\'emico
de la Universidad Nacional Aut\'onoma de M\'exico.
Support for this work was provided by European project FP7-ICT BrainScales (to G.D. and M.M.-G ).
In addition, G.D. was supported by the European Research Council  Advanced Grant: DYSTRUCTURE (n. 295129), by the Spanish Research Project SAF2010-16085.
A.T.C. was supported by the European Community's Seventh Framework Programme (FP7/2007-2013) under Grant Agreement PEOPLE-2012-IEF-329837.


\newpage

\begin{figure*} [ht] %
\centering
   \includegraphics[width=0.6\columnwidth]{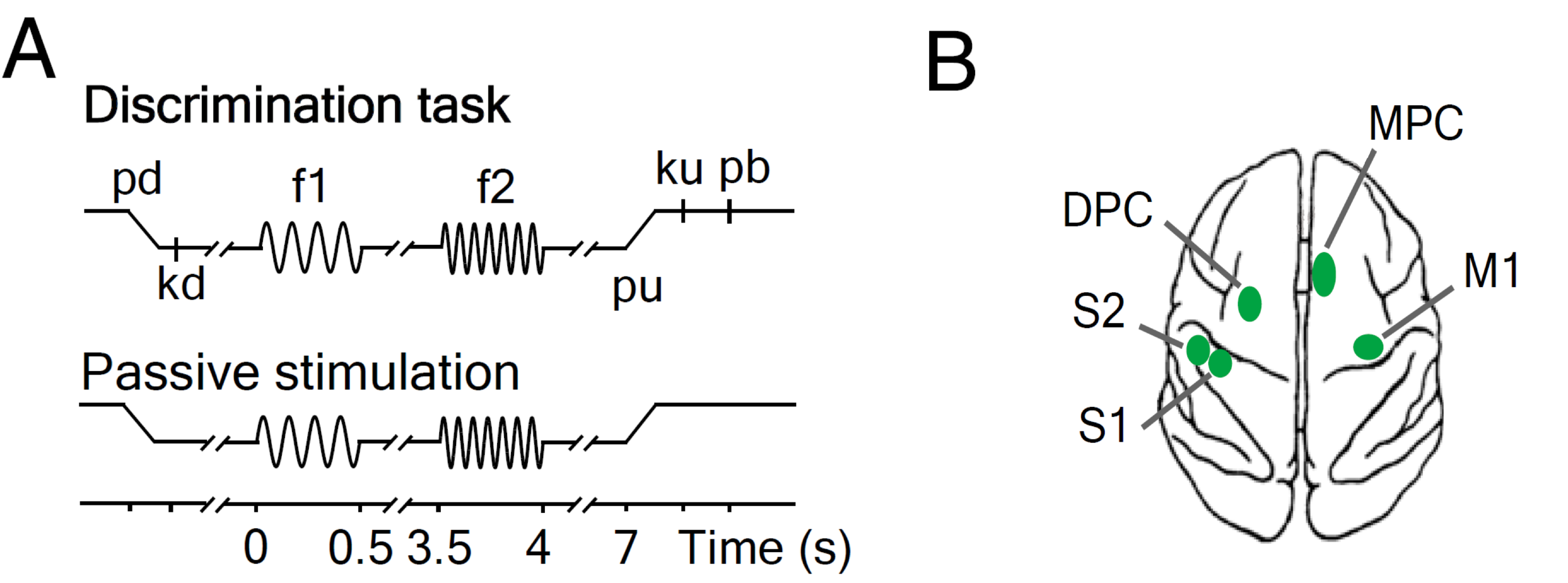}
   \caption{ 
 Somatosensory discrimination task and cortical recoding sites. (A) Sequence of events during the discrimination and the passive stimulation tasks ($f1$, first stimulus; $f2$, second stimulus; \emph{kd}, key down; \emph{ku}, key up; \emph{pb}, push button; \emph{pd}, probe down; \emph{pu}, probe up; 
 (\emph{Materials and Methods}). (B) Top view of the monkey brain and the recorded cortical areas (green spots).
   S1, primary somatosensory cortex; S2, secondary somatosensory cortex;  MPC, medial premotor cortex; DPC, dorsal premotor cortex; M1, primary motor cortex. }
\end{figure*}
\vspace{-5mm}

\begin{figure*} [h]
\centering
   \includegraphics[width=1.0\columnwidth]{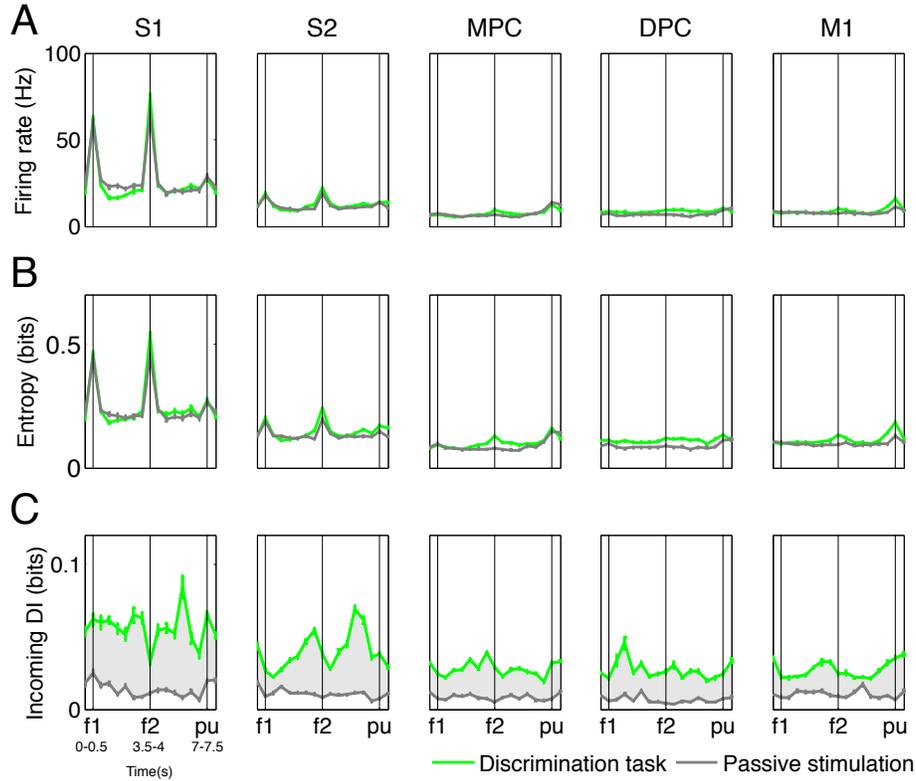}
   \caption 
   { 
   Single-neuron vs. multiple-neuron measures in the first monkey. 
     Comparison between discrimination (green) and passive stimulation tasks (grey) across areas using the average value of distinct measures over 
   the ensemble of neurons with incoming responsive paths.     Data was  obtained in $13$ sessions ($n=13$) from areas
   S1, primary somatosensory cortex; S2, secondary somatosensory cortex; DPC, dorsal premotor cortex; MPC, medial premotor cortex; M1, primary motor cortex, and is plotted for $17$ consecutive intervals when  $f1=14$Hz and $f2=22$Hz.       Vertical bars outline the intervals $f1$, $f2$ and \emph{pu} period.  Error bars ($\pm$ 2SEM) denote the standard deviation of the sample mean. 
  (A) Average firing rate. (B) Average entropy. (C) Average (across the ensemble of neurons) sum of directed information (DI) along incoming responsive paths.  
The shadowed grey area indicates the difference  of this measure between both tasks.}
    \vspace*{5mm}
\end{figure*}
   \begin{figure*}[h]
\centering
   \includegraphics[width=1.0\columnwidth]{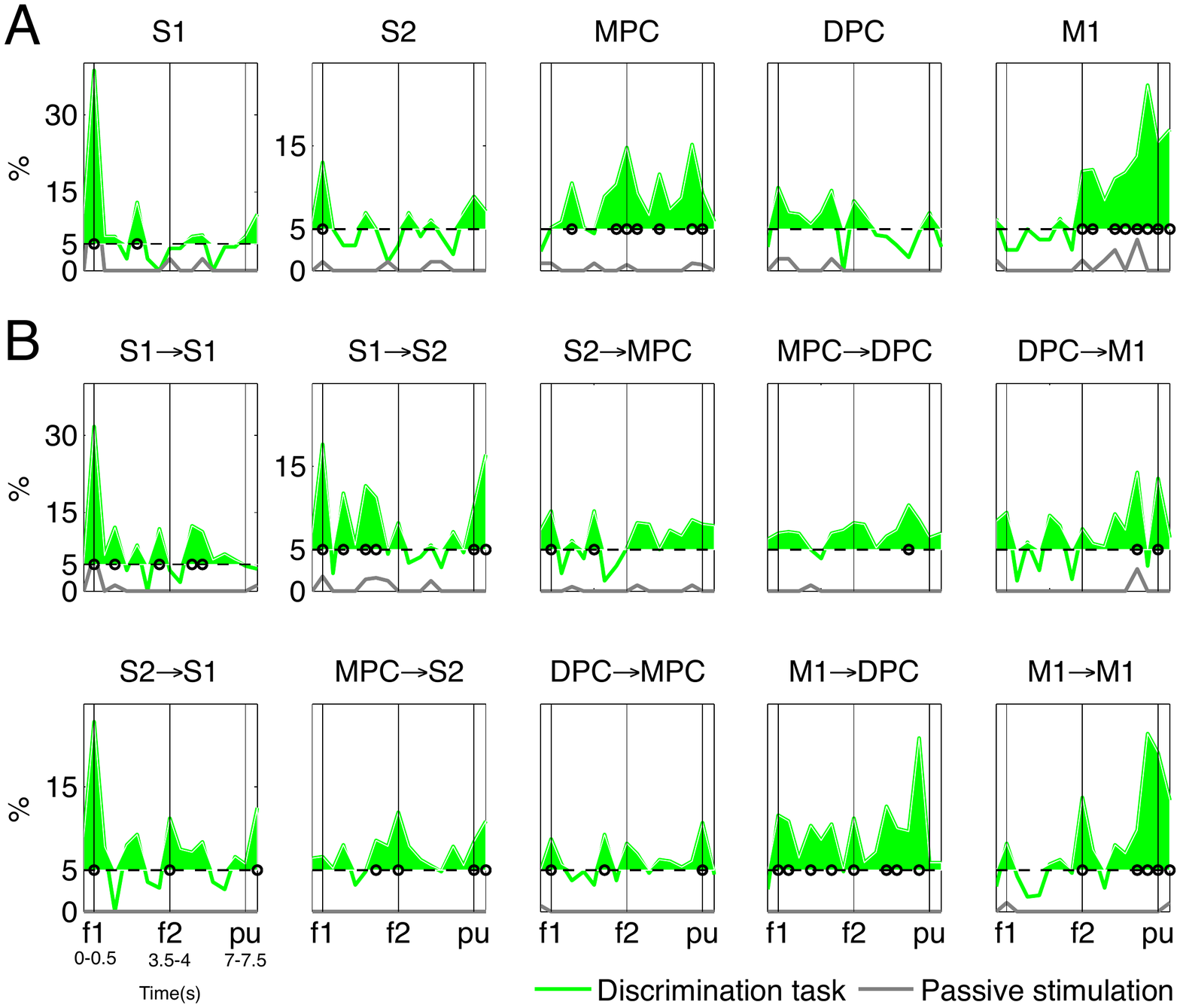}
   \caption 
   {  
   Modulated neurons and paths in the first monkey.
    In green, percentages  during the discrimination task.  In grey,  percentages during passive stimulation.
    Arrows in the title indicate the directionality of the modulated paths. 
   Vertical bars outline the intervals $f1$, $f2$ and \emph{pu} period.  
   Horizontal dashed lines indicate significance level $\alpha=5\%$.  
  The shadowed green area indicates the percentages of modulated paths above significance level.   
  Black circles indicate the intervals where the estimated percentage was significantly different (Agresti-Coull confidence interval \cite{Ag98}, $\alpha=5\%$)
from significance level.
   (A) Percentage of modulated neurons over all responsive neurons in each recorded area. 
   (B) Percentage of modulated paths over all responsive paths in $10$ intra- and interarea comparisons. 
           Data was  obtained in $13$ sessions ($n=13$) from areas
   S1, primary somatosensory cortex; S2, secondary somatosensory cortex; DPC, dorsal premotor cortex; MPC, medial premotor cortex; M1, primary motor cortex, and is plotted for $17$ consecutive intervals. 
}
\end{figure*}

\begin{figure*}[h]
\centering
   \includegraphics[width=1.0\columnwidth]{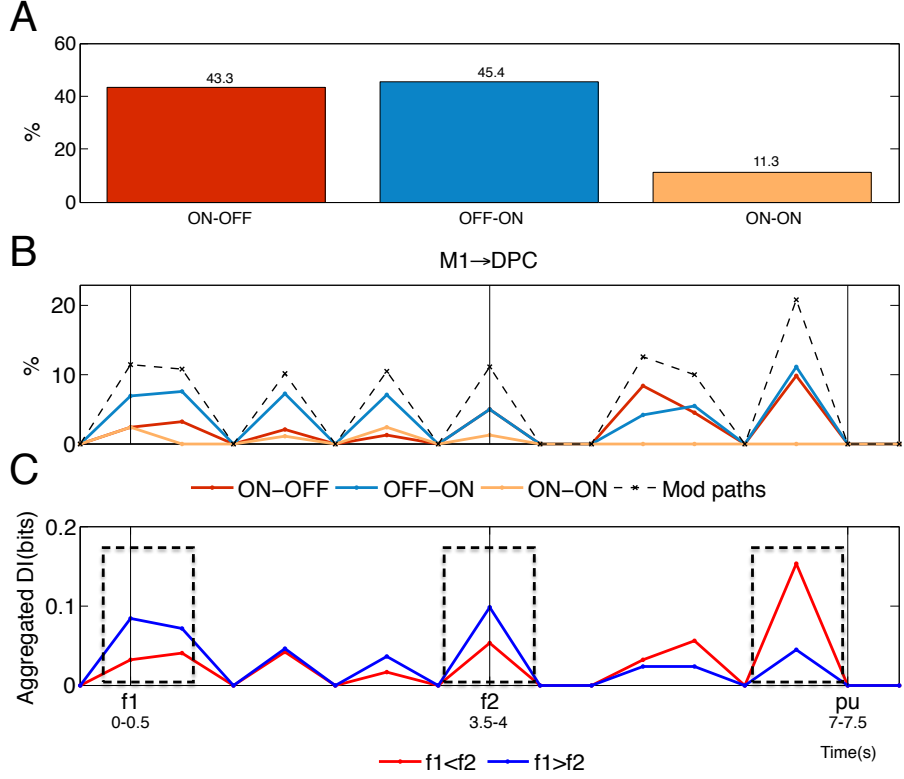}
   \caption 
   { 
   Modulation classes during the discrimination task in the first monkey.
    (A) Distribution of modulated paths from intervals above significant level ($\alpha=5\%$)  into the classes ON-OFF, OFF-ON and ON-ON. 
    (B)  For the comparison  M1-DPC, percentage of modulation classes during task intervals above significant level: 
percentages of ON-OFF modulations (significant only for $f1<f2$, red),  percentages of OFF-ON modulations (significant only for $f1>f2$, blue), 
and percentages of ON-ON modulations (significant for both decision reports, orange). For reference, the total percentage of modulated paths is 
plotted in a dashed black line with cross markers. 
(C) Aggregated sum  
of the average (across trials) directed information (DI) along modulated paths from M1 to DPC  during 
task intervals above significant level  for the decision reports  ($f1=14$Hz, $f2=22$Hz), ($f1<f2$, red) and   ($f1=30$Hz, $f2=22$Hz), ($f1>f2$, blue). (B)-(C)  
Arrows in the title indicate the directionality of the modulated paths. 
Vertical bars outline the intervals $f1$, $f2$ and \emph{pu} period.  Data was  obtained in $13$ sessions ($n=13$) from areas
   S1, primary somatosensory cortex; S2, secondary somatosensory cortex; DPC, dorsal premotor cortex; MPC, medial premotor cortex; M1, primary motor cortex, and is plotted in (B) and (C) for $17$ consecutive intervals.}
\end{figure*}
%

\begin{figure*}[h]
\centering
   \includegraphics[width=1.0\columnwidth, scale=0.32]{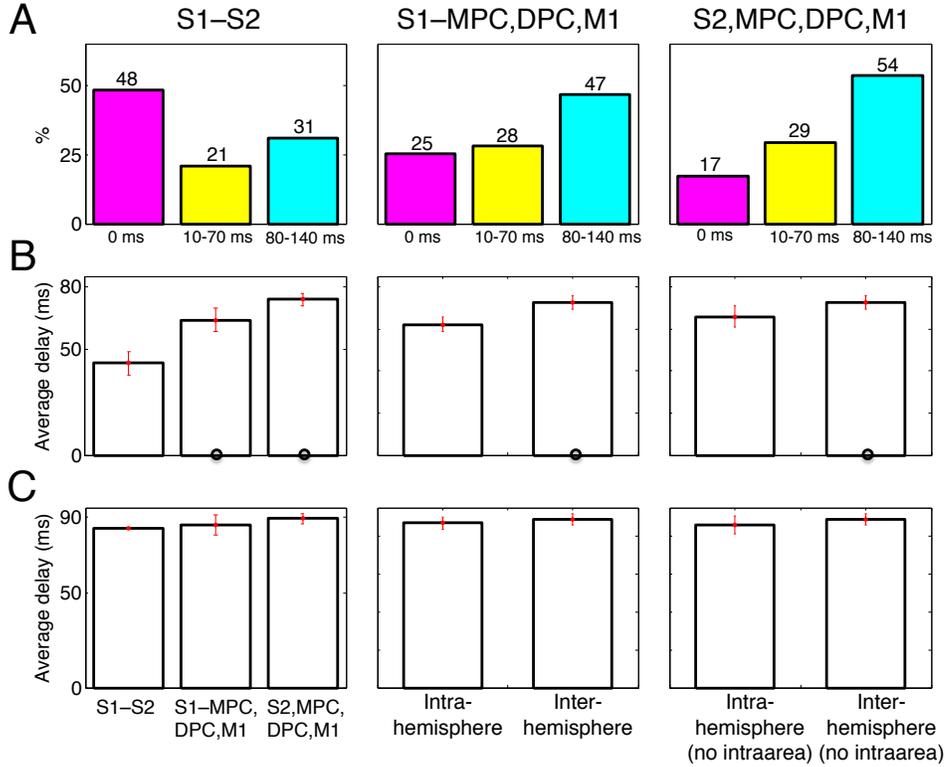}
   \caption 
   { 
   Interneuronal delays during the discrimination task in the first monkey.    (A) Distribution of modulated path delays at intervals above significant level ($\alpha=5\%$) into $0$ms (magenta),  
   $[10,70]$ms (yellow) and  $[80,140]$ms (cian) in correlations across S1 and S2 (left);
   correlations between S1 and MPC, DPC and M1 (center);  
   and correlations across S2, DPC, MPC and M1 (right). 
   (B)-(C) Average delay of correlations across  S1 and S2; correlations between S1 
   and MPC, DPC and M1; and correlations across S2, DPC, MPC and M1 (left). Average delay of correlations
   across areas within the same hemisphere (``intrahemisphere") and across areas from opposite hemispheres (``interhemisphere") (center). 
   Average delay of correlations
   across areas within the same hemisphere excluding  correlations within the same area (``intrahemisphere no intraarea") and across areas 
   from opposite hemispheres (right). Error bars in red ($\pm$ SEM)  denote the standard deviation of the delays. Black circles indicate  that the difference
   against the leftmost category was significant (Wilcoxon test, $\alpha=5\%$). 
   (B) Average delay of  modulated paths. 
   (C) Average delay of non-instantaneous modulated paths.   
   Data was  obtained in $13$ sessions ($n=13$) from areas
   S1, primary somatosensory cortex; S2, secondary somatosensory cortex; MPC, medial premotor cortex;  DPC, dorsal premotor cortex; M1, primary motor cortex.
}
\end{figure*}

\end{document}


\maketitle
\tableofcontents

 \newpage
 \section{Supplementary figures}

 \begin{figure}[!htbp]\centering
  \includegraphics[width=1\columnwidth]{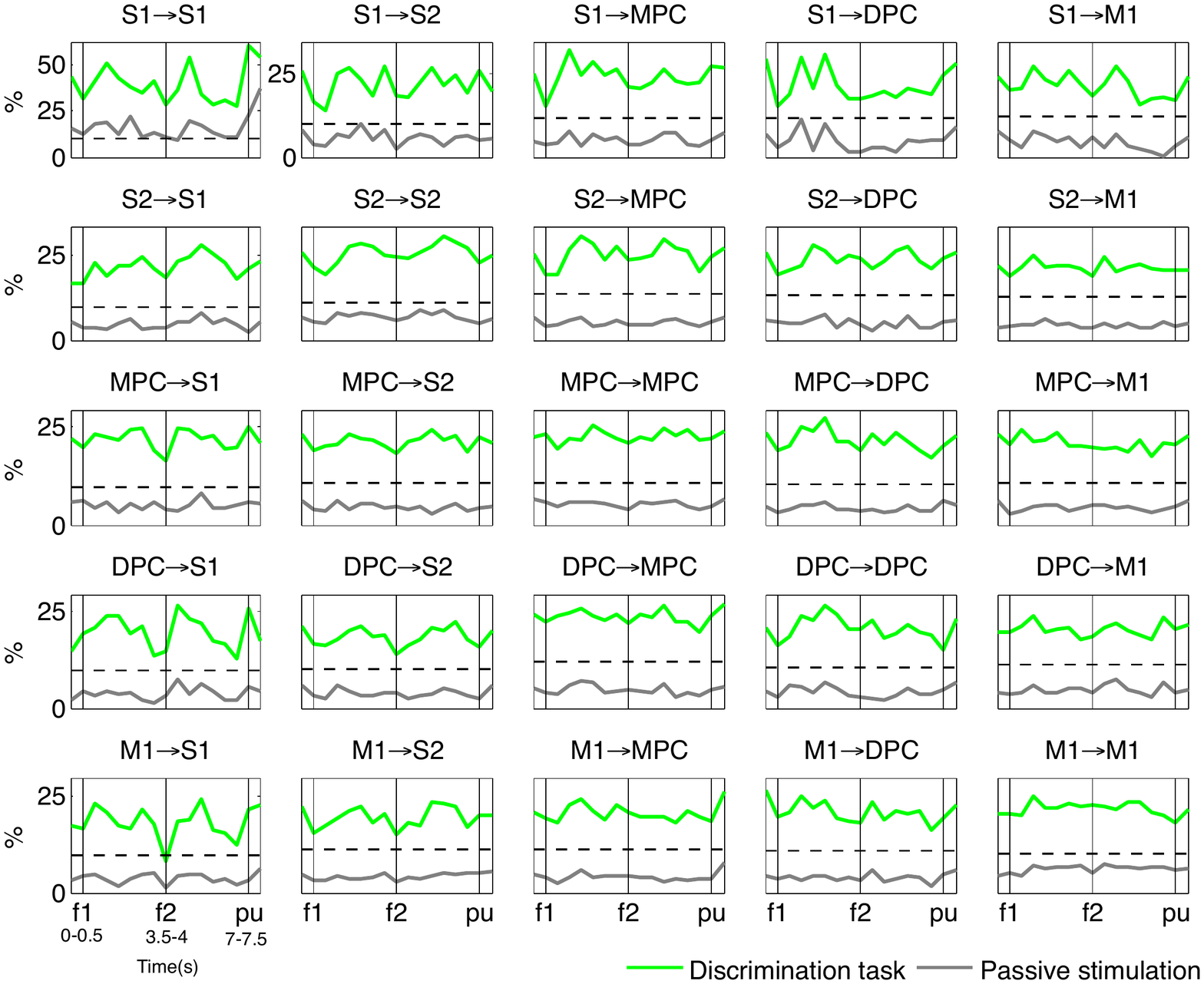}
  \caption{Responsive paths in the first monkey. 
  Percentage of responsive paths in all interarea comparisons during $17$ consecutive task intervals. 
   Arrows in the title indicate the directionality of the modulated paths.  Vertical bars outline the intervals $f1$, $f2$ and \emph{pu} period.  Horizontal dashed lines indicate significance level ($\alpha'=9.75\%$, where $\alpha'=2\alpha(1-\alpha)+\alpha^{2}$ and $\alpha=5\%$). In green, percentages of responsive paths during the discrimination task.  In grey,  percentages of responsive paths 
   whose correlations were also significant for either the frequency pair ($f1=14$Hz, $f2=22$Hz) or ($f1=30$Hz, $f2=22$Hz) during passive stimulation. 
   Data were  obtained in $13$ sessions ($n=13$) from areas
   S1, primary somatosensory cortex; S2, secondary somatosensory cortex; MPC, medial premotor cortex;  DPC, dorsal premotor cortex; M1, primary motor cortex,
    and were plotted for $17$ consecutive  intervals.  }
  \end{figure}

\begin{figure}[!htbp]\centering
  \includegraphics[width=1\columnwidth]{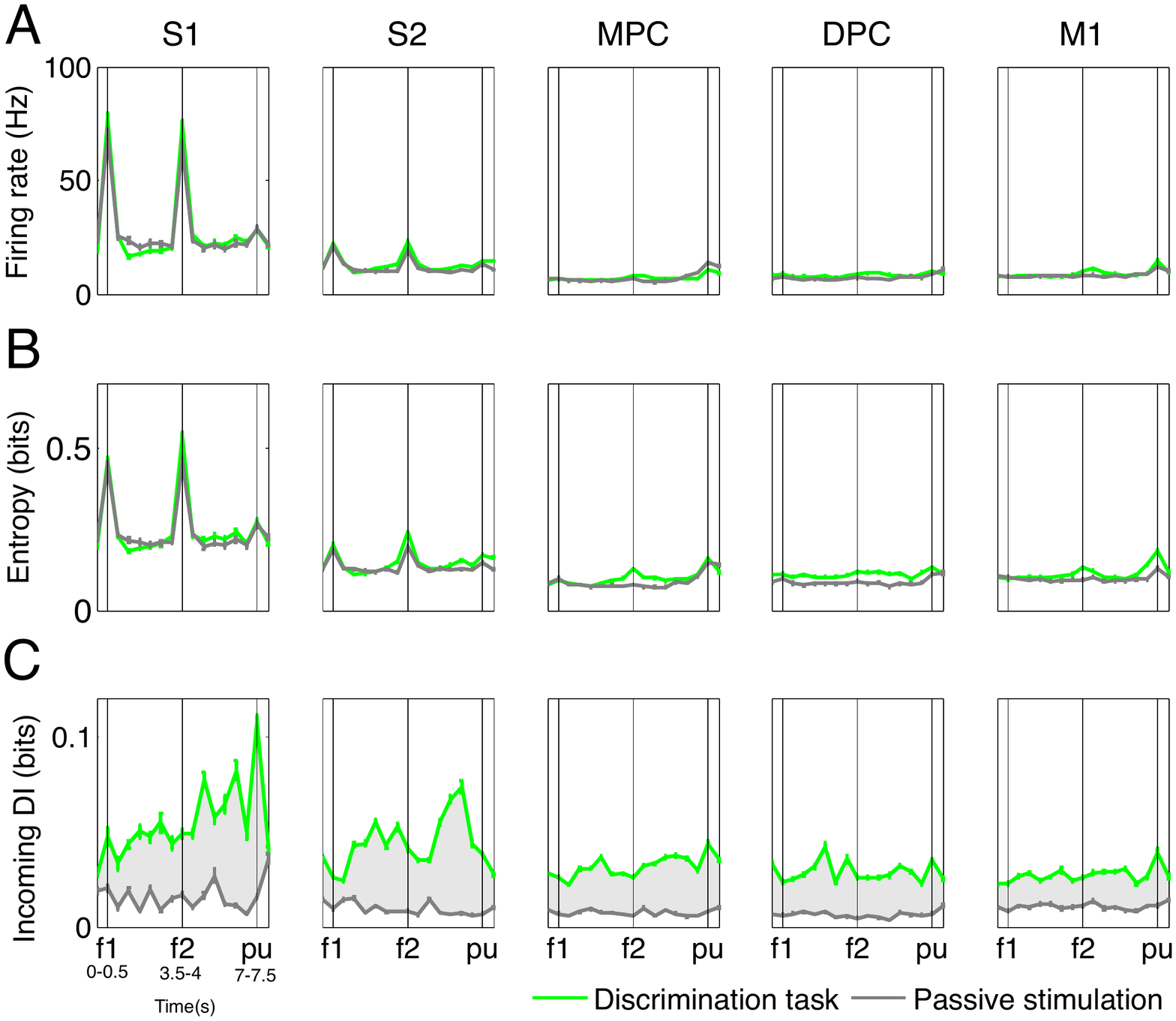}
  \caption{Single-neuron vs. multiple-neuron measures in the first monkey. Comparison between discrimination (green) and passive stimulation tasks (grey) across areas using the average value of distinct measures over the ensemble of neurons with incoming responsive paths. 
   Vertical bars outline the intervals $f1$, $f2$ and \emph{pu} period. 
        Data were  obtained in $13$ sessions ($n=13$) from areas
   S1, primary somatosensory cortex; S2, secondary somatosensory cortex; MPC, medial premotor cortex;  DPC, dorsal premotor cortex; M1, primary motor cortex,
    and were plotted for $17$ consecutive  intervals
 when $f1=30$Hz and $f2=22$Hz. 
  Error bars ($\pm$ SEM) denote the standard error of each measure.
  (A) Average firing rate. (B) Average entropy. (C) Average (across the ensemble of neurons) sum
  of directed information along incoming  responsive paths. The shadowed grey area indicates the difference  of this measure between both tasks.}
  \end{figure}

\begin{figure}[!htbp]\centering
  \includegraphics[width=1\columnwidth]{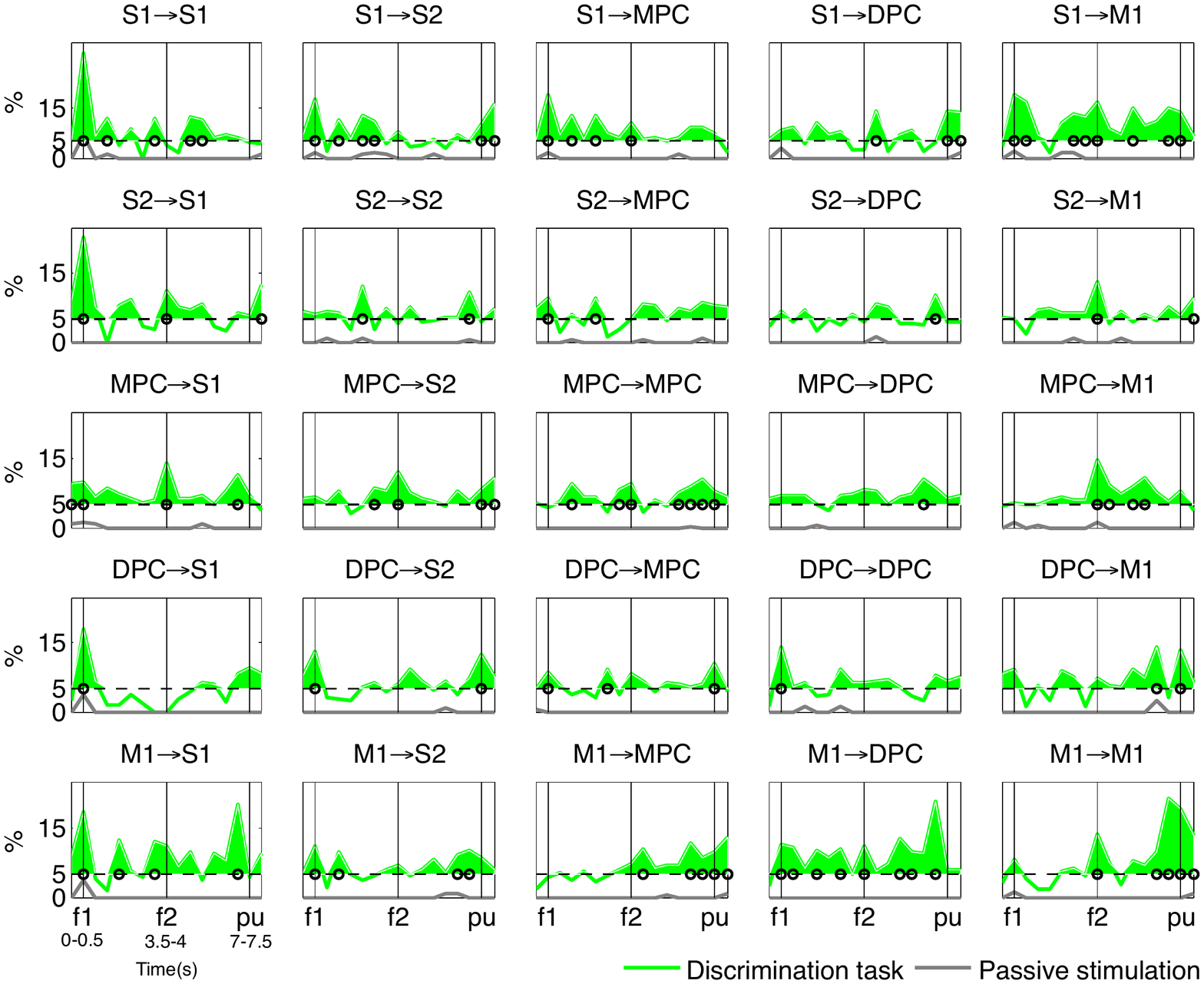}
  \caption{ Modulated paths in the first monkey. Percentage of modulated paths over responsive paths in all  
  intra- and interarea comparisons during $17$ consecutive task intervals. 
 In green, percentages  during the discrimination task.  
   In grey,  percentages during passive stimulation.  
   Arrows in the title indicate the directionality of the modulated paths.  Vertical bars outline the intervals $f1$, $f2$ and \emph{pu} period.  
   Horizontal dashed lines indicate the significance level ($\alpha=5\%$). The shadowed green area indicates the percentages of modulated paths above significance level. 
    Black circles indicate the intervals where the estimated percentage was significantly different (Agresti-Coull confidence interval \cite{Ag98}, $\alpha=5\%$)
from significance level.        Data were  obtained in $13$ sessions ($n=13$) from areas
   S1, primary somatosensory cortex; S2, secondary somatosensory cortex; MPC, medial premotor cortex;  DPC, dorsal premotor cortex; M1, primary motor cortex,
    and were plotted for $17$ consecutive  intervals. 
}
   \end{figure}
  
  \begin{figure}[!htbp]\centering
  \includegraphics[width=1\columnwidth]{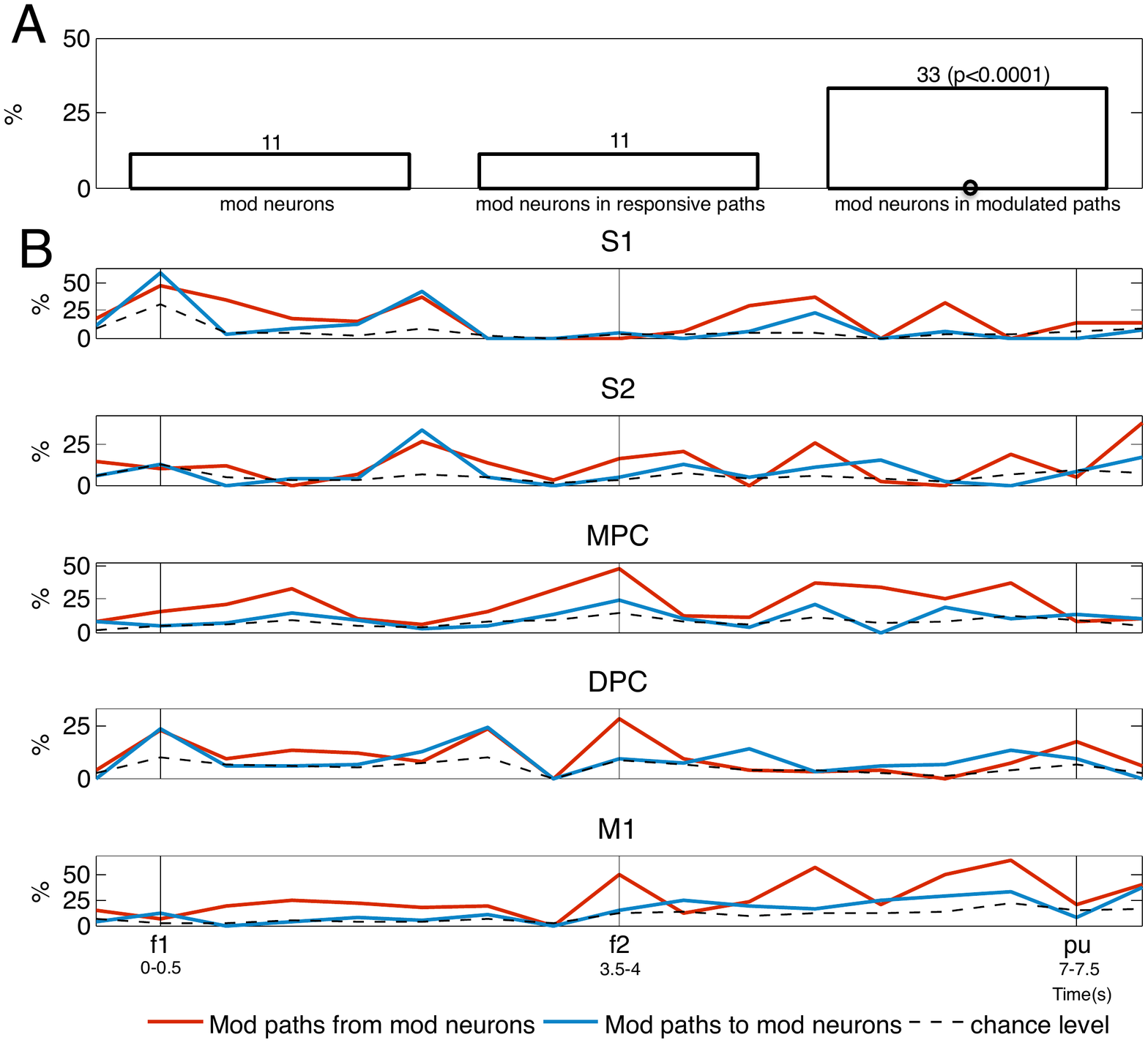}
  \caption{Relationship between modulated neurons and modulated paths in the first monkey. 
  (A) Comparison of the proportion of modulated neurons in all tested neuron pairs (``mod neurons"), responsive (``mod neurons in responsive paths" ) and modulated paths 
  (``mod neurons in modulated paths" ). The black circle highlights that there was a significant correlation between modulated neuron and the existence of an own outgoing or incoming modulated path.
 (B) Proportion
  of modulated paths whose starting point neuron or endpoint was a modulated neuron in each recorded area. 
  In red, percentage of outgoing modulated paths from modulated neurons over all modulated outgoing paths from an area (``Mod paths from mod neurons'"). 
  In blue, percentage of incoming modulated paths to modulated neurons over all modulated incoming paths to an area (``Mod paths to mod neurons'"). 
  In dashed black, probability that a modulated neuron was the starting point or endpoint neuron of a randomly selected neuron pair (`chance level"). 
   Vertical bars outline the intervals $f1$, $f2$ and \emph{pu} period. 
  Data were  obtained in $13$ sessions ($n=13$) from areas
   S1, primary somatosensory cortex; S2, secondary somatosensory cortex; MPC, medial premotor cortex;  DPC, dorsal premotor cortex; M1, primary motor cortex,
    and were plotted for $17$ consecutive  intervals.  }
 \end{figure}

\begin{figure}[!htbp]\centering
  \includegraphics[width=1\columnwidth]{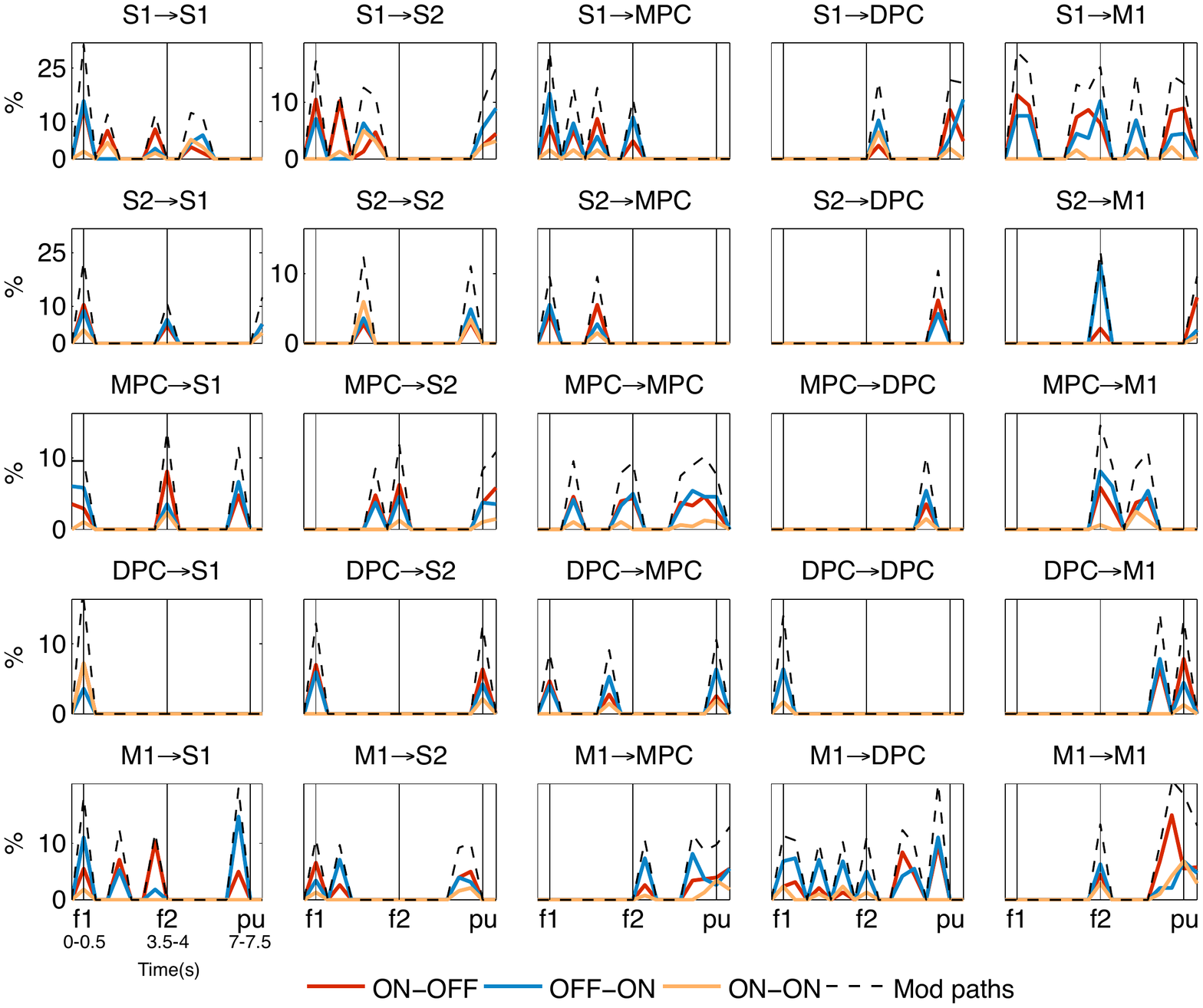}
  \caption{ Modulation classes during the discrimination task in the first monkey. Percentage of modulation types  in all
   interarea comparisons and task intervals above significant level ($\alpha=5\%$):
percentages of ON-OFF modulations (significant only for $f1<f2$, red),  OFF-ON modulations (significant only for $f1>f2$, blue), 
and ON-ON modulations (significant for both, orange). For reference, the total percentage of modulated paths were plotted in a dashed black line.
Arrows in the title indicate the directionality of the modulated paths. Vertical bars outline the intervals $f1$, $f2$ and \emph{pu} period. 
  Data were  obtained in $13$ sessions ($n=13$) from areas
   S1, primary somatosensory cortex; S2, secondary somatosensory cortex; MPC, medial premotor cortex;  DPC, dorsal premotor cortex; M1, primary motor cortex,
    and were plotted for $17$ consecutive  intervals. 
  }
   \end{figure}

  \begin{figure}[!htbp]\centering
    \includegraphics[width=1\columnwidth]{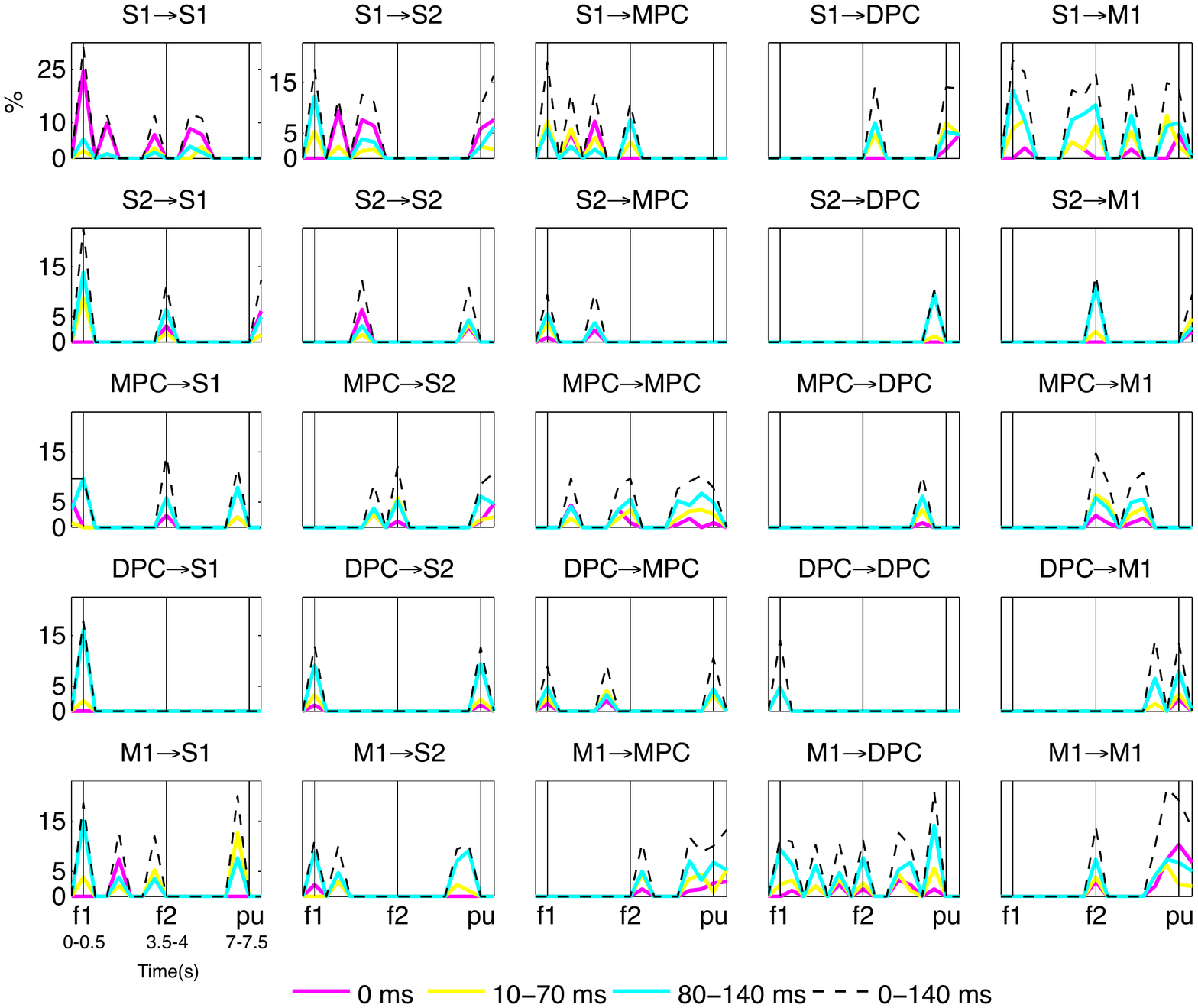}
  \caption{Modulated path delays during the discrimination task in the first monkey. 
  Percentage of modulated path delays in all
   interarea comparisons and task intervals above significant level ($\alpha=5\%$):
 percentages of instantaneous correlations ($0$ms, magenta), percentage of modulated paths at  delays within $10-70$ms (yellow) and percentages of 
 modulated paths at delays within  $80-140$ms (cian).  For reference, the total percentage of modulated paths were plotted in a dashed black line. Arrows in the title indicate the directionality of the modulated paths.  Vertical bars outline the intervals $f1$, $f2$ and \emph{pu} period. 
  Data were  obtained in $13$ sessions ($n=13$) from areas
   S1, primary somatosensory cortex; S2, secondary somatosensory cortex; MPC, medial premotor cortex;  DPC, dorsal premotor cortex; M1, primary motor cortex,
    and were plotted for $17$ consecutive  intervals. 
 }
   \end{figure}

\begin{figure}[!htbp]\centering
  \includegraphics[width=1\columnwidth]{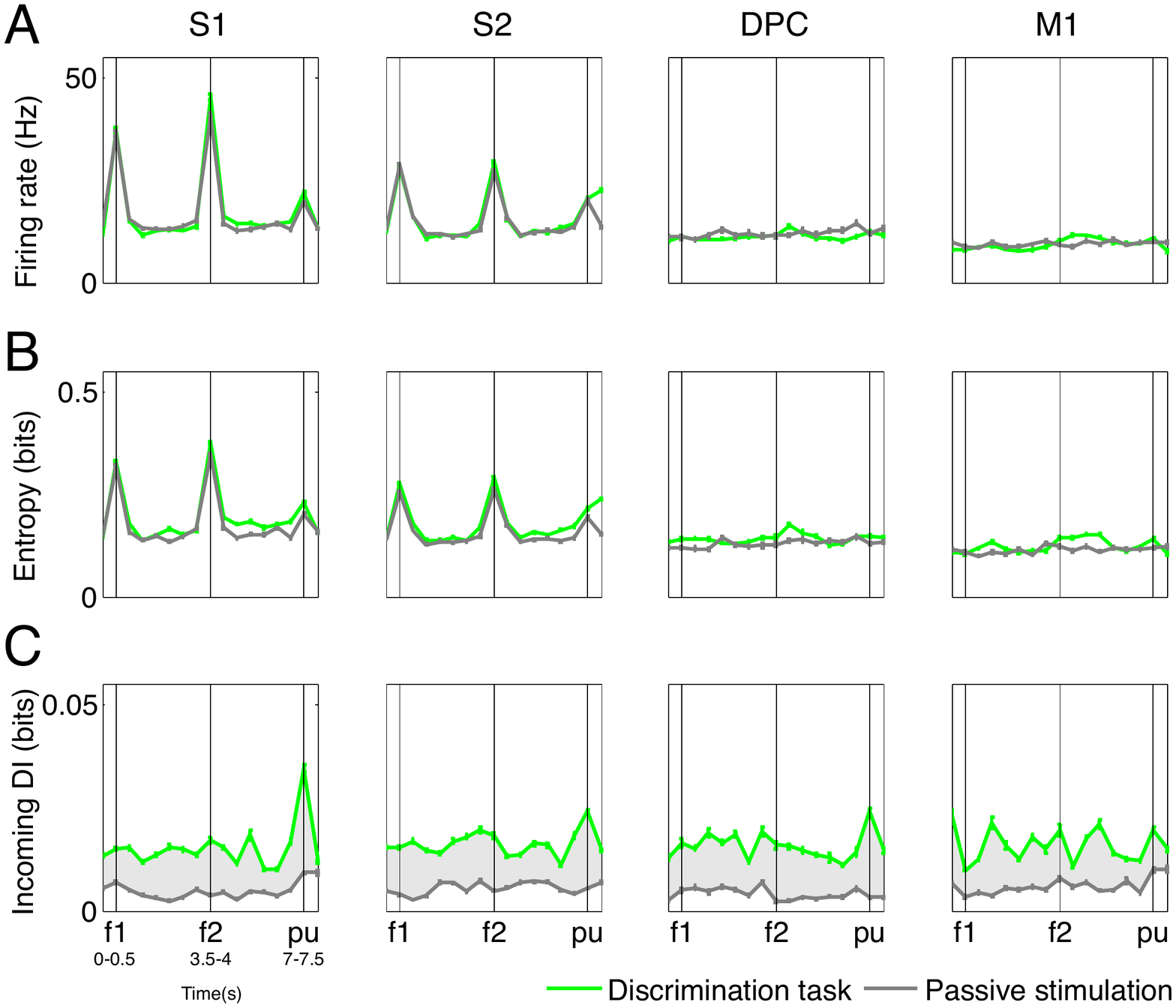}
  \caption{Single-neuron vs. multiple-neuron measures in the second monkey. Comparison between discrimination (green) and passive stimulation tasks (grey) across four areas using the average value of distinct measures over the ensemble of neurons with incoming responsive paths. 
        Data were  obtained in $19$ sessions ($n=19$) from areas
S1, primary somatosensory cortex; S2, secondary somatosensory cortex;  DPC, dorsal premotor cortex;  and   S1, primary somatosensory cortex; S2, secondary somatosensory cortex;  and M1, primary motor cortex and were plotted for $17$ consecutive  intervals
 when $f1=14$Hz and $f2=22$Hz. 
   Vertical bars outline the intervals $f1$, $f2$ and \emph{pu} period. Error bars ($\pm$ 2SEM) denote the standard error of each measure.
  (A) Average firing rate. (B) Average entropy. (C) Average (across the ensemble of neurons) sum
  of directed information along incoming  responsive paths. The shadowed grey area indicates the difference  of this measure between both tasks.}
  \end{figure}


\begin{figure}[!htbp]\centering
  \includegraphics[width=1\columnwidth]{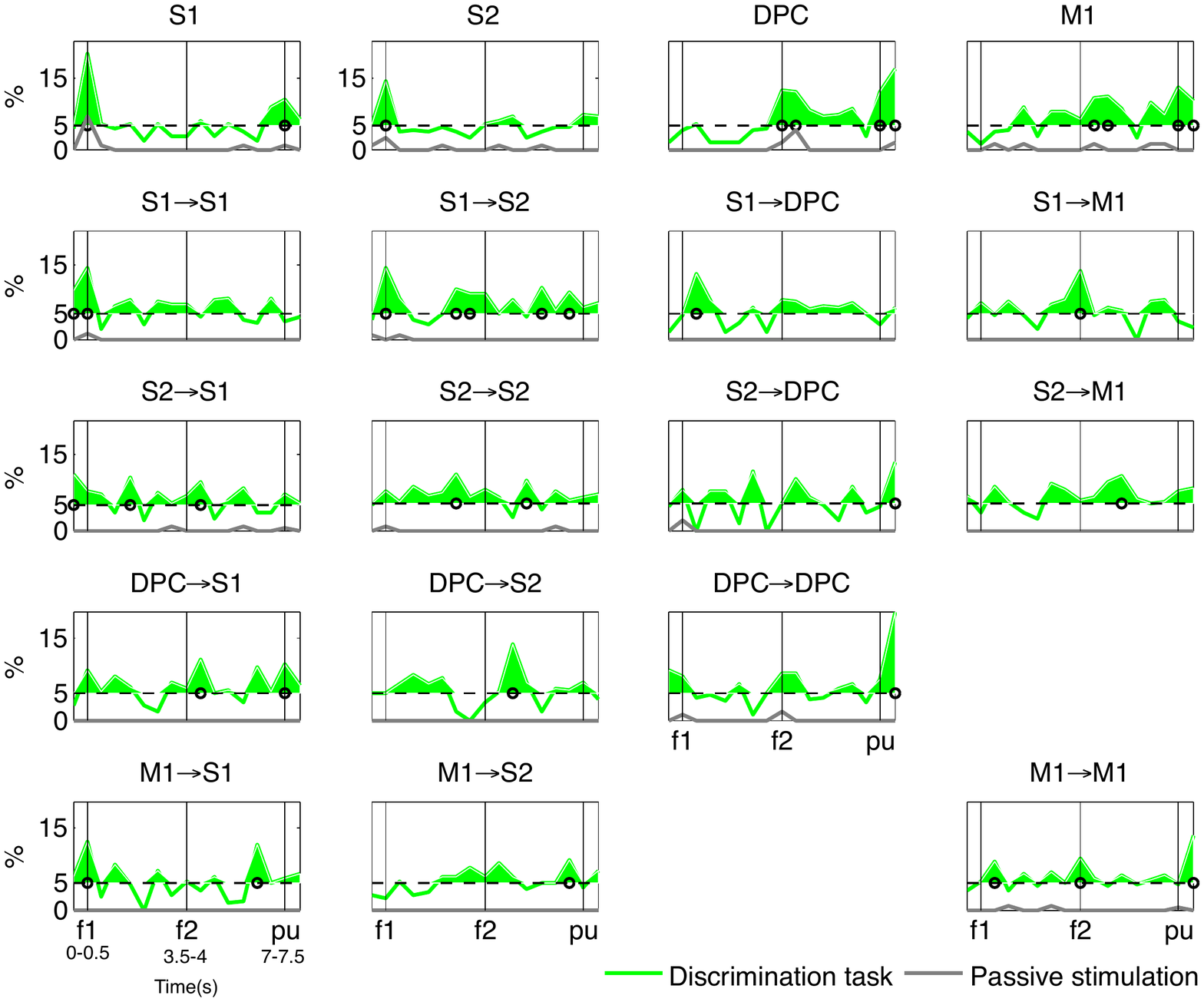}
  \caption{  Modulated neurons and paths in the second monkey.
    In green, percentages  during the discrimination task.  In grey,  percentages during passive stimulation.
    Arrows in the title indicate the directionality of the modulated paths. 
   Vertical bars outline the intervals $f1$, $f2$ and \emph{pu} period.  
   Horizontal dashed lines indicate significance level $\alpha=5\%$.  
  The shadowed green area indicates the percentages of modulated paths above significance level.   
  Black circles indicate the intervals where the estimated percentage was significantly different (Agresti-Coull confidence interval \cite{Ag98}, $\alpha=5\%$)
from significance level.
   (A) Percentage of modulated neurons over all responsive neurons in each recorded area. 
   (B) Percentage of modulated paths over all responsive paths in $10$ intra- and interarea comparisons. 
           Data were  obtained in $19$ sessions ($n=19$) from 
  simultaneous  areas S1, primary somatosensory cortex; S2, secondary somatosensory cortex;  DPC, dorsal premotor cortex;  and   S1, primary somatosensory cortex; S2, secondary somatosensory cortex;  and M1, primary motor cortex, and
were plotted for $17$ consecutive  intervals.}
 \end{figure}

  \begin{figure}[!htbp]\centering
  \includegraphics[width=1\columnwidth]{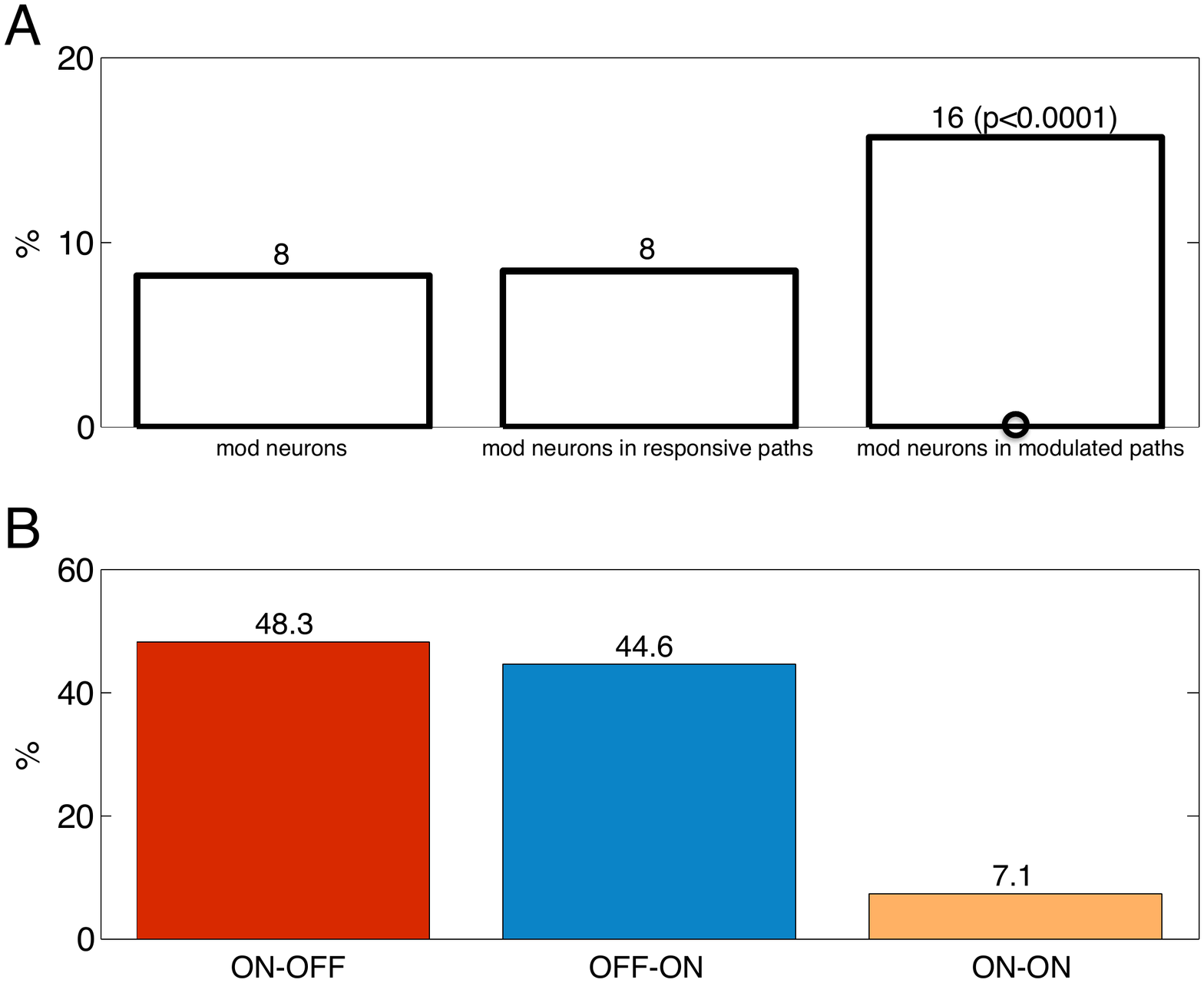}
  \caption{Additional results for the second monkey.
  (A) Relationship between modulated neurons and modulated paths. Comparison of the proportion of modulated neurons in all tested neuron pairs (``mod neurons"), responsive (``mod neurons in responsive paths" ) and modulated paths 
  (``mod neurons in modulated paths" ). The black circle highlights that there was a significant correlation between modulated neuron and the existence of an own outgoing or incoming modulated path. 
    (B)  Modulation classes during the discrimination task.
    Distribution of modulated paths from intervals above significant level ($\alpha=5\%$)  into the classes ON-OFF, OFF-ON and ON-ON. 
  }
 \end{figure}

\newpage

\section{Glossary of terms}
\begin{itemize}
\item Path: non-linear and (possibly) delayed directional correlation between two neurons. In general, there is no direction 
defined over a path, but it has an starting point (influencing neuron) and an 
endpoint neuron (influenced neuron). In this work, correlations are computed using  the directed information measure \cite{Mas90}. 
\item Incoming path (to a neuron): a path  whose endpoint is the neuron under consideration. 
\item Outgoing  path (from a neuron): a path whose starting point is the neuron under consideration. 
\item Responsive neuron: a neuron with significant entropy (permutation test, $\alpha = 5\%$) for
at least one frequency pair. 
\item Responsive path: a path between responsive neurons for which the value of the directed information (permutation test, $\alpha = 5\%$) 
is significant  for
at least one frequency pair. 
\item Modulated neuron: a responsive neuron with significant differences (permutation test, $\alpha = 5\%$) in its entropy 
between the sets of trials $(f1 = 14, f2 = 22)$Hz and ($f1 = 30, f2 = 22)$Hz.
\item Modulated path: a responsive path with significant differences (permutation test, $\alpha = 5\%$) in the value of the directed information  
between the sets of trials $(f1 = 14, f2 = 22)$Hz and ($f1 = 30, f2 = 22)$Hz.
\item  ON-ON modulated path: modulated path with significant directed information for both frequency pairs, $(f1 = 14, f2 = 22)$Hz and $(f1 = 30, f2 = 22)$Hz.
\item ON-OFF modulated path: modulated path with significant directed information for
the frequency pair $(f1 = 14, f2 = 22)$Hz  
 but non-significant for the frequency pair $(f1 = 30, f2 = 22)$Hz.
\item  OFF-ON modulated path: modulated path with significant directed information for
the frequency pair $(f1 = 30, f2 = 22)$Hz  
 but non-significant for the frequency pair $(f1 = 14, f2 = 22)$Hz.
\end{itemize}

\section{Estimation of the directed information}\label{sec:estimation}

\subsection{Notation}

Let 
$X^T=(X_1,\dotsc, X_T)$ and $Y^T=(Y_1,\dotsc, Y_T)$ be two random processes that describe the time series $x^T=(x_1, \dotsc, x_T)$ and $y^T=(y_1, \dotsc, y_T)$.
 We shall use $X_i$ to denote the $i$-th component of $X^T$ and $X^i_j=(X_i,\dotsc, X_j)$, $i<j$, to denote a subset of consecutive components of $X^T$.
We shall denote the distribution of
the joint process $(X^T, Y^T)$ as $P_{X^T Y^T}$
with marginal distributions  $P_{X^T}$ and $P_{Y^T}$. 



\subsection{Introduction}
The majority of methods that estimate information-theoretic quantities between  two 
random processes $X^T$ and $Y^T$ 
are based on the computation of the underlying joint probability distribution of a presumed
jointly ergodic and stationary process $(\mathcal{X},\mathcal{Y})$.  
A commonly used estimator in computational neuroscience is the \emph{plug-in} estimator, which 
estimates the underlying joint  distribution by tracking the frequency of string occurrences in an observed time series \cite{Bes10,Panzeri+2010}. 
The main drawback of this estimator is the undersampling problem: since all strings are assumed to be equally likely, the estimator 
requires a sufficiently large number of trials to ensure convergence.
Nonetheless,  some bias reduction techniques have been
proposed to increase the convergence of this estimator \cite{Panzeri+2007,Bes10}.
In this work, we follow a Bayesian approach based on the context-tree weighting (CTW) algorithm,  \cite{Will95, Jiao13}, which 
	has been proved to outperform the bias and the variance of the plug-in estimator\footnote{An exhaustive study of the performance differences between the plug-in and the CTW estimator can be found in \cite{Gao+CTW+2008}. }. 

In the next sections we provide a general overview of the CTW method. 
Further implementation details as well as properties of this method can be found in \cite{Will95}.
We start by introducing the concept of tree source model upon which the algorithm is built.



\subsection{Tree source model}

 
We consider that sequences of a $M$-ary alphabet (in our case $M$=2) are generated by a \emph{tree source} of bounded memory $D$, which means 
that the
generation of a symbol $x_t$  depends on a suffix of its most recent $D$ symbols $x^{t-1}_{t-D}$. 
More formally stated,
the probability of the generated sequence is defined by the model $(\Scal,\Theta_{\Scal})$, where $\Scal$ is the \emph{suffix set}
consisting of $M$-ary strings of length no longer than $D$, and
\begin{equation}
\Theta_{\Scal}=(\boldsymbol{\theta}_s; s\in \Scal)
\end{equation}
is the parameter space  where $\boldsymbol{\theta}_s\triangleq (\theta_{0,s}, \theta_{1,s}, \dotsc, \theta_{M-2, s})$. The suffix set is required to be 
\emph{proper} (suffixes in the set are not suffixes of other elements of $\Scal$) and \emph{complete} (every sequence has a suffix in $\Scal$). 
Then, we can define a mapping $\beta_{ \Scal}(\cdot)$ by which every recent $D$ symbols, $x^{t-1}_{t-D}$, are mapped to a unique suffix $s\in \Scal$. To each suffix, there corresponds 
a parameter vector $\boldsymbol{\theta}_s$ that determines the next symbol probability in the sequence as

\begin{equation}
\Pr\left\{X_t=i| x^{t-1}_{t-D}, \Scal, \Theta_{ \Scal}\right\}=\theta_{i,\beta_{ \Scal}(x^{t-1}_{t-D})}
\end{equation}
for $i=0,\dotsc,  M-2$, and 
\begin{equation}
\Pr\left\{X_t=M-1| x^{t-1}_{t-D}, \Scal, \Theta_{ \Scal}\right\}=1-\prod_{i=0}^{M-2}\theta_{i,\beta_{ \Scal}(x^{t-1}_{t-D})}.
\end{equation}

The goal of the algorithm is to estimate the
probability of any sequence generated by a tree source without knowing the underlying model $(\Scal, \Theta_{ \Scal})$, i.e, without knowing neither 
the suffix set $\Scal$ nor the parameter space $\Theta$. 

\paragraph*{Example:}
Let $M=2$, $D=2$ and consider the suffix set $\Scal=\{00, 10,1\}$. Then, the probability of the sequence $x_{1}^{7}=0110100$, where $x_1=0, x_2=1,\dots,x_7=0$ given 
the past symbols $10$ can be evaluated as $\Pr\left\{x_1^7|\Scal, \theta_{00}, \theta_{10}, \theta_1\right\}$: 
\begin{align*}
\Pr(0110100 | 10)=&P(0|10)  \cdot  P(1|00) \cdot  P(1|01) \cdot  P(0|11) \cdot  P(1|10)  \cdot  P(0|01) \cdot  P(0|10)\\
                                                           =&(1-\theta_{10}) \cdot  \, \theta_{00}  \, \cdot  \,  \theta_1  \cdot (1-\theta_1) \cdot  \, \,\theta_{10} \cdot (1-\theta_1)  \cdot (1-\theta_{10}),
 \end{align*}
where we used the mapping  
 $\beta_{\Scal}(10)=10$, $\beta_{\Scal}(00)=00$,
 $\beta_{\Scal}(01)=1$ (the sufix 01 is not in the set of suffixes $\Scal$, and we thus map it to the suffix one  $\beta_{\Scal}(11)=1$.

\subsection{Bayesian approach}

The context-tree weighting is a method of approximating the true probability of a $T$-length sequence $x_1^T$ generated according to the true model $(\Scal^{\star}, 
\boldsymbol{\theta}^{\star})$  with the mixture probability
\begin{equation}\label{mix_prob}
\hat{P}(x_1^T)_=\sum_{(\Scal, \Theta_{\Scal})} w(\Scal,\Theta_{\Scal}) P_{\Scal, \Theta_{\Scal}} (x_1^T),
\end{equation}
where $w(\cdot)$ is a weighting function over all tree models and $P_{\Scal, \Theta_{ \Scal}}(x_1^T)$ is the probability of generating 
the sequence $x_1^T$ according to the model $(\Scal, \Theta_{\Scal})$. 

To approximate \eqref{mix_prob}, we first make use of the concept of context tree. The context tree is a set of nodes where each node is an $M$-ary
string $s$ with length $l(s)$, and where $l(s)$ is upper-bounded by a given memory $D$. Each node $s$ splits into $M$ (child) nodes $0s, 1s, \dotsc, (M-1)s$. 
To each node there corresponds a vector of counts $\a_s=(a_{0,s},a_{1,s},\dotsc, a_{M-1, s})$ of the number of times that a symbol is preceded by the string $s$. For a parent node $s$ and its children $0s, 1s, \dotsc, (M-1)s$,
the counts must satisfy $a_{i,s}=\sum_{j=0}^{M-1} a_{i, js}$ for every symbol $i=0,\dotsc, M-1$. Then, for every node with string $s$ we estimate the probability
that a sequence is generated with the counts $\a_s$.
 Counts in each node are updated by each new observation $x_t$, $t=1, \dotsc, T$.  
%

In general, the probability that a memoryless source with parameter vector $\boldsymbol{\theta}=(\theta_1,\theta_2,\dots, \theta_M)$ generates a given sequence follows a multinomial 
distribution. By averaging this probability over all possible values of $\theta_i$, $i=1,\dots,M$, with a Dirichlet distribution we obtain the Krichevsky-Trofimov (KT) probability estimator. 
A useful property of this estimator is that it can be sequentially computed as $P^s_e(0,0, \dotsc, 0)=1$ and
\begin{equation}
 P^s_e(a_{0,s},a_{1,s},\dotsc, a_{i-1,s}, a_{i,s}+1,  a_{i-1,s},\dotsc, a_{M-1,s})=\frac{a_{i,s}+\frac{1}{2}}{a_{0,s}+a_{1,s}+\dotsc+ a_{M-1,s}+\frac{M}{2}}.
\end{equation}

Finally, we assign a probability to each node, which  is the weighted combination of the estimated probability and the weighted probability of its children:
\begin{align}
P_{w}^{s}=\begin{cases}
P_w^s=\alpha P_e^s(\a_s)+(1-\alpha)\prod_{i=1}^MP_w^{is}, \quad & 0\leq l(s)<D\\
P_e^s(\a_s), \quad  &l(s)=D,
\end{cases}
\end{align}
where $\alpha$ is typically chosen to be $\frac{1}{2}$.

\subsection{Schematic version of the algorithm for an $M-$ary alphabet}\label{section:algorithm}

For every $t=1,\dotsc, T$, we use the context $x^{t-1}_{t-D}$ and the value of $x_t$. Then, we track nodes from the leaf to the root node along
the path determined by $x^{t-1}_{t-D}$.

\begin{itemize}

\item \textbf{Leafs}: Identify the leaf $s$ that corresponds to  $x^{t-1}_{t-D}$ in the context tree. Then
\begin{enumerate}
\item \emph{Counts update}\\ 
Based on the value of $x_t$,  update 
$\a_s$.
\item \emph{Estimated probability}\\
Compute $P_e^s(\a_s)$ using the Krichevsky-Trofimov estimator, which is defined recursively as $P^s_e(0,0\dotsc 0)=1$ and for $a_{i,s}\geq 0$, $i=1, \cdots,M-1$,
\begin{equation*}
 P_e^s(a_{0,s},a_{1,s},\dotsc, a_{i-1,s}, a_{i,s}+1,  a_{i-1,s},\dotsc, a_{M-1,s})=\frac{a_{i,s}+\frac{1}{2}}{a_{0,s}+a_{1,s}+\dotsc+ a_{M-1,s}+\frac{M}{2}}.
\end{equation*}

\item \emph{Weighted probability}\\
For the leaf nodes, $P_w^s=P_e^s(\a_s)$. 
\end{enumerate}

\item \textbf{Internal nodes}: Using the path determined by the context $x^{t-1}_{t-D}$,\\
REPEAT
\begin{enumerate}
\item \emph{Parent search}\\
Identify the  parent $s$ of the previously tracked node.
\item \emph{Counts update}\\ 
Based on the value of $x_t$, update 
$\a_s$.

\item \emph{Estimated probability}\\
Compute $P_e^s(\a_s)$ using $\a_s$ and the Krichevsky-Trofimov estimator.

\item \emph{Weighted probability}\\
Compute $P_w^s$ as 
\begin{equation*}
P_w^s=\alpha P_e^s(a_{0,s},a_{1,s},\dotsc, a_{M-1,s})+(1-\alpha)\prod_{i=1}^MP_w^{is},
\end{equation*}
where $\alpha$ is typically chosen to be $\frac{1}{2}$.\\
\end{enumerate}
UNTIL the root node is tracked. 

\item \textbf{Probability assignment}: Let $\lambda$ denote  the root node of the context tree. Then, $\hat{P}(x^t)\equiv P_w^{\lambda}(x^n)$ 
is the universal probability assignment in the CTW algorithm. As a result, we also obtain the conditional probability $\hat{P}(x_1^t|x_1^{t-1})$ as:
\begin{equation*}
\hat{P}(x_1^t|x_1^{t-1})=\frac{P_w^{\lambda}(x_1^{t})}{ P_w^{\lambda}(x^{t-1}_1)}.
\end{equation*}
\end{itemize}

\begin{figure} [!htbp]
         \begin{center}
       \includegraphics[width=0.3\columnwidth]{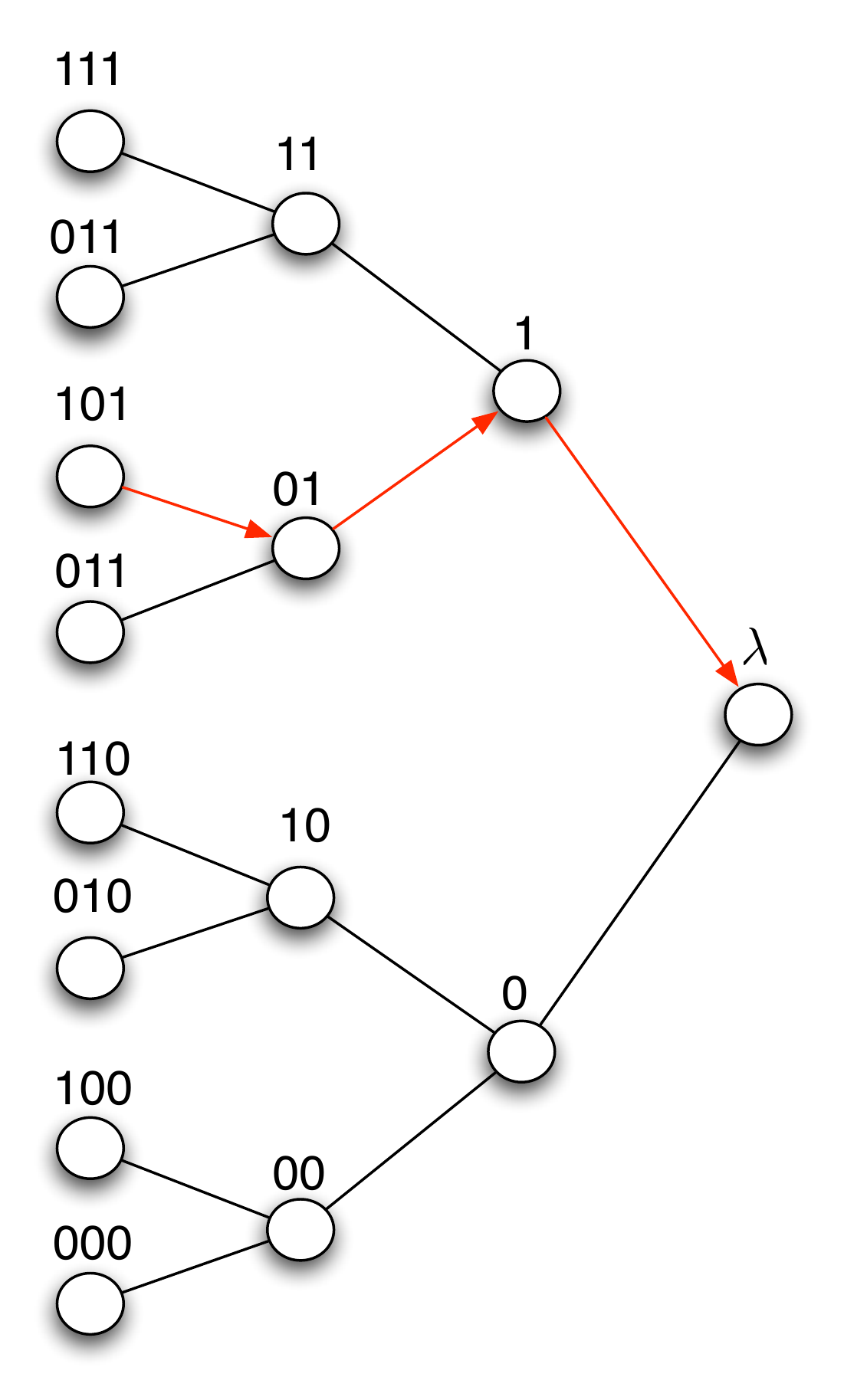}
          \caption{Context tree with the path determined by the context $x^{0}_{-2}=101$ in red.}
           \label{CTW1}
          \end{center}
\end{figure}

\begin{figure} [!htbp]
         \begin{center}
       \includegraphics[width=0.5\columnwidth]{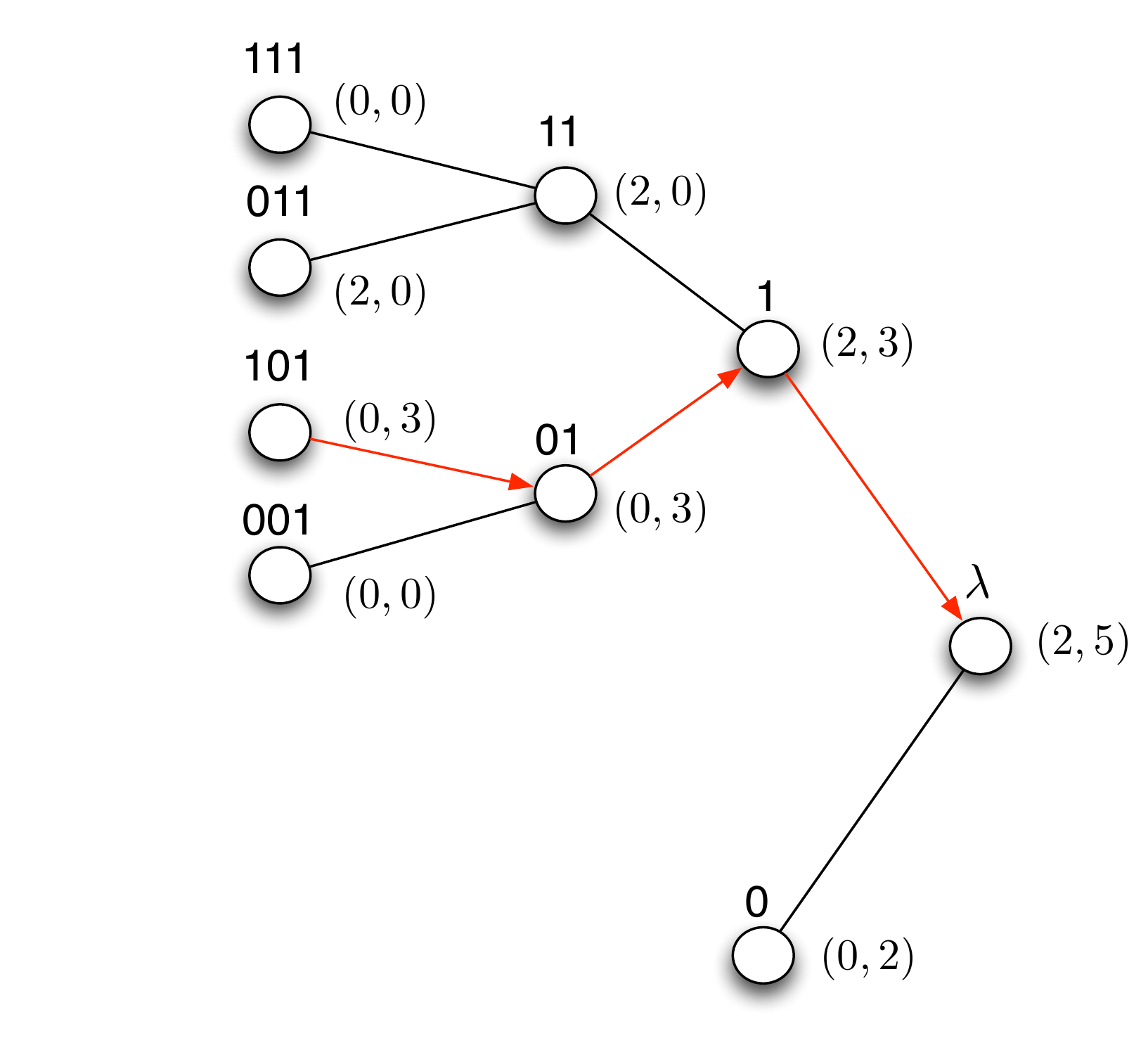}
          \caption{Counts update up to $x_7=1$. }
           \label{CTW2}
          \end{center}
\end{figure}

\begin{figure} [!htbp]
         \begin{center}
       \includegraphics[width=1.0\columnwidth]{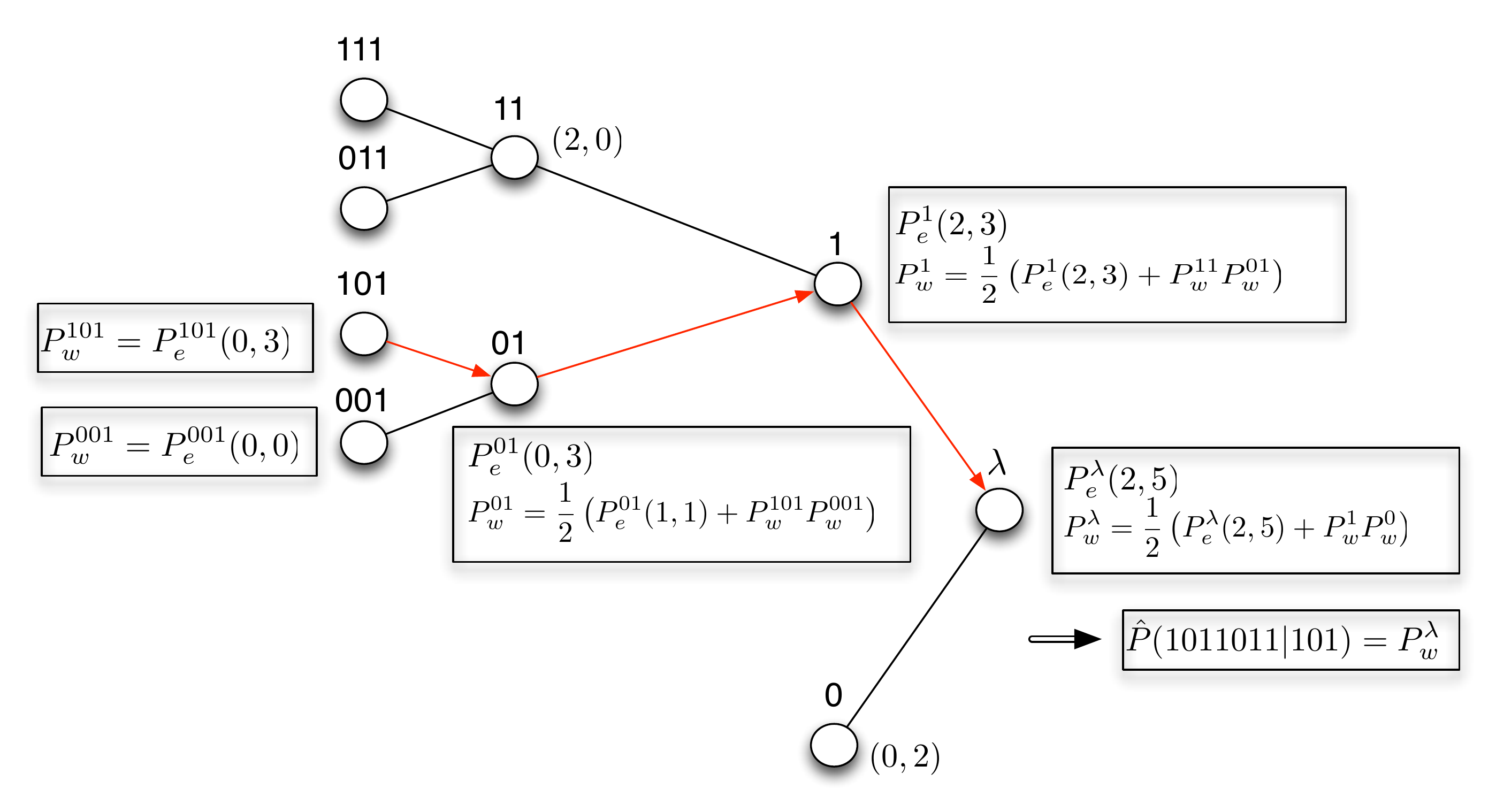}
          \caption{Backward computation of estimated, $P_e^s(\a_s)$, weighted probabilities, $P_w^s$, at  $x_7=1$,  and probability assignment, $\hat{P}(x_1^7)$ .}
           \label{CTW3}
          \end{center}
\end{figure}

\paragraph*{Example:}
Consider the binary  sequence $x^7=1011011$ with past symbols $x_{-2}^{0}=101$. 
We evaluate the context tree for $M=2$ and $D=3$. Suppose that we are at time instance $t=1$ where the context is $101$ (Fig. \ref{CTW1}).
After observing the sequence up to $t=7$, we obtain counts $\a_s=(a_{0,s}, a_{1,s})$ for each context tree node  (Fig. \ref{CTW2}). 
From the leafs to the root node ($\lambda$), we recursively compute the weighting probabilities 
and provide the probability assignment $\hat{P}(x^7)$ (Fig. \ref{CTW3}).

\newpage

\subsection{Estimator based on the CTW algorithm}\label{DI_estimation}
The estimator of the directed information that we employ  is built upon the CTW algorithm \cite{Jiao13}.
Then, given a simultaneous observation $(x^T,y^T)$,  we must assume that it is a realization of a jointly stationary finite-alphabet Markov chain $(\mathcal{X}, \mathcal{Y})$ with memory $D$ to ensure estimation consistency. The formula to compute the estimator is the following: 
 \begin{align}
  \hat{I}(\mathcal{X}\rightarrow \mathcal{Y})\triangleq 
 & \frac{1}{T}\sum_{t=1}^T \sum_{y_t} \hat{P}(Y_t=y_t\bigl |X^t_{t-D}=x^t_{t-D}, Y^{t-1}_{t-D}=y^{t-1}_{t-D})\notag\\
 & \times \log \frac{\hat{P}(Y_t=y_t \bigl| X^t_{t-D}=x^t_{t-D}, Y^{t-1}_{t-D}=y^{t-1}_{t-D})}{\hat{P}(Y_t=y_t\bigl|Y^{t-1}_{t-D}=y^{t-1}_{t-D})},\label{eq1}
  \end{align}
  where the probabilities are estimated using the context-tree weighting method. 
We next summarize the main steps of this computation:
\begin{enumerate}
\item Estimation of  the probabilities $\hat{P}(Y_t=y_t \bigl|Y^{t-1}_{t-D}=y^{t-1}_{t-D})$ and $\hat{P}(X_t=x_t, Y_t=y_t \bigl| X^{t-1}_{t-D}=x_{t-D}^{t-1} , Y^{t-1}_{t-D}=y_{t-D}^{t-1})$.
%
\item Computation of the marginal probability 
\begin{equation}\label{eq4}
\hat{P}\bigr(X_t=x_t \bigl|X^{t-1}_{t-D}=x^{t-1}_{t-D}, Y^{t-1}_{t-D}=y^{t-1}_{t-D} \bigl)=\sum_{y_t} \hat{P}\bigl(X_t=x_t,Y_t=y_t \bigl | X^{t-1}_{t-D}=x^{t-1}_{t-D}, Y^{t-1}_{t-D}=y^{t-1}_{t-D}\bigr).
\end{equation}

\item  Application of Bayes theorem using \eqref{eq4}: 
\begin{equation}\label{eq5}
\hat{P}\bigr(Y_t=y_t \bigl| X^t_{t-D}=x^t_{t-D}, Y^{t-1}_{t-D}=y^{t-1}_{t-D} \bigl)=\frac{P\bigl(X_t=x_t, Y_t=y_t \bigl| X^{t-1}_{t-D}=x^{t-1}_{t-D}, Y^{t-1}_{t-D}=y^{t-1}_{t-D}\bigr)}{\hat{P}\bigr(X_t=x_t \bigl|X^{t-1}_{t-D}=x^{t-1}_{t-D}, Y^{t-1}_{t-D}=y^{t-1}_{t-D} \bigl)}.
\end{equation}

\item Plug-in of  \eqref{eq5} and $\hat{P}(Y_t=y_t \bigl|Y^{t-1}_{t-D}=y^{t-1}_{t-D})$ into \eqref{eq1} to obtain $\hat{I}(X^T\rightarrow Y^T)$.

\end{enumerate}

 \section{Data preprocessing} 

  \subsection{Preliminary selection of neurons}
   We selected $n=13$ recorded sessions from one monkey and $n=19$ recorded sessions from a second monkey. 
   In Tables S1 and S2 
   we summarize the selected neurons per area and session in the discrimination and passive task.

     \begin{table} [!htbp] \centering     
      \begin{tabular}{l|cccccc}\label{tab:tab1}
    Session/Area  & S1  & S2  & MPC & DPC & M1&  \\ \hline
     1 & 5   &  8  & 13  &  4  &  8 & \\
    2 & 6   &  7  & 12  &  9  &  9 & \\
    3 &   5 & 12  &  13 &  9  &  6 & \\
    4 &  5  & 4   & 11  &  8  &  5 & \\ 
    5 &  1 & 9   & 15  &  3  &  5 & \\ 
    
    6 & 7   &  7  & 10  &  5  &  6 & \\
    7 & 2   &  16  & 2  &  6  &  6 & \\
    8 &  2 & 1   & 16  &  2  &  7 & \\ 
    
    9 & 1   &  11  & 11  &  4  &  8 & \\
    10 & 0   &  8  & 13  &  9  &  5 & \\
    11 &  5 & 2  &  13 &  4  &  5 & \\
    12 &  4  & 8   & 7  &  6  &  10 & \\ 
    13 &  4 & 9   & 10  &  6  &  8 & \\ 
    \hline \hline 
    TOTAL & 47 & 102  &  146 &  75 & 88 & 458\\ 
    \end{tabular}
    \caption[Number of neurons per area and session]{Number of neurons per area and session from monkey 1.}
    \end{table}
    
         \begin{table} [!htbp] \centering     
      \begin{tabular}{l|cccccc}\label{tab:tab1}
    Session/Area  & S1  & S2  & DPC & M1&  \\ \hline
     1 & 4   &  10  & 0  &  5     \\
    2 & 5   &  8  & 0  &  9     \\
    3 & 7 & 10  &  0 &  8     \\
    4 &  4  & 5   & 0  &  12     \\ 
    5 &  8 & 13   & 0  &  12     \\ 
    
    6 & 7   &  10 & 0  &  14     \\
    7 & 6  &  13  & 0  &  15     \\
    8 &  5 & 7   & 0  &  10     \\ 
    
    9 & 5   &  5  & 3  &  0     \\
    10 & 8  &  6  & 7 &  0     \\
    11 &  5 & 11  &  3 &  0     \\
    12 &  5 & 7   & 11  &  0    \\ 
    13 & 5  &  3  & 4  &  0     \\
    14 &  5 & 6  &  4 &  0     \\
    15 &  9  & 5   & 7  &  0    \\ 
    16 &  4 & 2   & 5  &  0     \\ 
    17 &  9 & 8   & 13  &  0   \\ 
    18 &  6 & 6   & 12  &  0   \\ 
    19 &  8 & 1   & 7  &  0   \\ 
    \hline \hline 
    TOTAL & 115 & 136  &  76 & 85 & 412\\ 
    \end{tabular}
    \caption[Number of neurons per area and session]{Number of neurons per area and session from monkey 2.}
    \end{table}

%
%
%
    
    For each session, we analyzed the following frequency pairs:
    \begin{equation*}
    \Bigl\{(f1=14,f2=22)\text{Hz},(f1=30,f2=22)\text{Hz}\Bigr\}.
    \end{equation*}
    We chose the pairs according to two criteria. The first criterion was to maintain the distance between the frequency pairs constant ($|f1-f2|=8$) to neglect
    effects due to the task difficulty. The second was to keep $f2$ fixed so that we were able to identify neural correlates of the decision after $f2$ stimulation.  
    We only used correct trials in the discrimination task.

    \subsection{Considerations about the estimator on spike-train data}\label{sec:application}
    
\begin{figure}[!htbp] \centering
      \includegraphics[width=0.9\columnwidth]{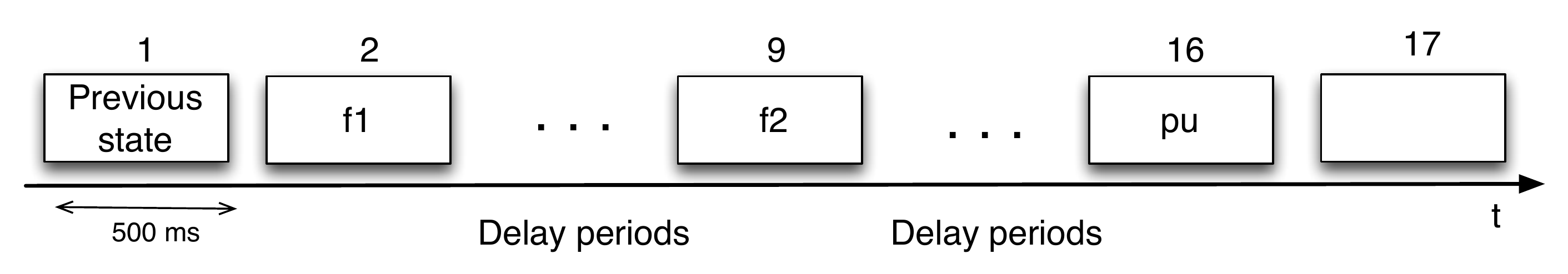}
      \caption[Trial division]{Schematic representation of the division of a trial of $8.5$s into $17$ intervals of $500$ms. 
      The second interval corresponds to the first stimulation, the ninth interval corresponds to the second stimulation 
       and the sixteenth interval corresponds to the probe-up period.}
     \label{fig:task_periods} 
     \end{figure}

   As introduced before, the consistency of the estimator  requires that any pair of simultaneously observed time series is a realization
 of a jointly stationary irreducible aperiodic Markov process of some bounded order. 
   However,  interactions between simultaneously recorded neural responses may occur at different delays  depending on the area and the task interval. Furthermore,   these interactions may be generated by statistically different processes. 
   To tackle these issues
%
we make the following assumptions:
   \begin{enumerate}
  \item  Spike trains can be binarized (i.e., assigning the value $1$ to each bin with at least one spike 
  and the value $0$, otherwise) using a bin size of $2$ms with limited information loss. This assumption is discussed is section \ref{ref:binning}. 
  
    \item Interactions occur 
    at interneuronal  delay values within the range $[0,140]$ms, which is chosen 
    based on the reaction times of each area \cite{Lafuente+romo+2006}.
    This range is binned into the sequence of delays $\delta=[0:5:70]$, i.e., $\delta=0(0\text{ms}) $, $5(10 \text{ms})$, $10(20 \text{ms})$, $\dots, 
   $70$ ($140$ \text{ms})$. We assume that interactions span $4$ms ($D=2$ bins) as it is suggested by a partial analysis of spike-trains entropies discussed in section \ref{sec:memory_delays}.  
%

  \item  
  We partition the task timeline 
  into $17$ consecutive task intervals of $500$ms, where two intervals match the stimulation periods (Fig. \ref{fig:task_periods}). Then, for each task interval and given delay  $\delta=[0:5:70]$ bins, any  pair of binarized spike trains $(x^{T-\delta}, y^T_{\delta+1})$, ($T=250$ bins) 
  satisfy the estimator conditions  with bounded memory $D=2$ bins. 
 \item  The underlying stationary process of each pair $(x^{T-\delta}, y^T_{\delta+1})$  is invariant across all trials recorded under
  the same frequency pair. 
 \end{enumerate}
 
    \subsubsection{Binarization of spike-train trials}\label{ref:binning}
   
   We evaluated the goodness of our bin choice by counting the number of times that more than one spike occurred in one bin and it was neglected. 
   The results  illustrated in Table S3 (for a sample of $5$ sessions with trials of $8$s, $n=5$) show that the number of  losses was at most $2.7$ spikes per trial. 
     \renewcommand{\arraystretch}{2} 
     \begin{table} [!htbp] \centering   
      \begin{tabular}{|c|c|c|c|c|c|}  
      \hline
          Area & S1  & S2  & MPC & DPC & M1  \\ \hline 
         Mean &   2.7  &  0.7&    0.1&    0.02&    0.083 \\ \hline 
    \end{tabular}
    \caption[Number of ISIs less than 2 ms and ratio of neglected spikes and ]{Average number of spike losses per trial ($8$s)
 in a sample of $5$ sessions recorded for the frequency pair $(f1=14,f2=22)$Hz. }
    \end{table}\label{tab:negleted_spikes}

    \subsubsection{Memory and delays}\label{sec:memory_delays}
   
  As introduced before, the performance of the CTW algorithm 
  depends on the maximum depth used, $D$, which can be interpreted as the memory of the Markov process underlying an observed time series. 
  Indeed, the computational cost of the algorithm grows exponentially with $D$, and $D$ therefore becomes a critical parameter to set when the number of required estimations is large.
  To obtain an approximation of neuronal memory 
  we calculated the entropy, $H(Y^T)$, of all neurons in one session for values of
  $D$ ranging from $0$ to $9$ during representative task intervals. 
  After inspecting how the  average entropy in each area under study stabilized as a function of the spike-train memory, 
  we chose a memory of $D=2$ bins($4$ms) as a good tradeoff between our empirical observation and the dimensionality of the parameter space that we wanted to estimate.

%

     A central question in our study is the time scale at which interactions occur. Results on interarea delays during decision
     making are scarce in the literature. Instead, the concept of task latency, i.e., the average time before an area is modulated by a task, 
     has been used
     to approximate the computation of delays during the whole discrimination task \cite{Lafuente+romo+2006}. Based on these results,
    we set the delays within the range $[0,140]$ms.      

 \section{Statistical procedures}
 
 Statistical tests were applied in two stages. First, we 
 computed significant values of the directed information across neuron pairs that were 
 simultaneously recorded to find responsive paths. 
 Then, we tested the modulation of significant correlations with respect to the monkey's decision report to find modulated paths.
 

      \subsection{Neuron-pair estimators}
    We first defined two estimators that were used to correct for multiple
    testing (one per delay) in each ordered neuron pair.  The two estimators were
    \begin{align}
 \hat{I}^{(1)}_{\Delta}(X^T\rightarrow Y^T)  \triangleq  \max_{\delta=[0:5:70]  }\hat{I}_{\delta}(\mathcal{X} \rightarrow \mathcal{Y}) \label{max_estimator}\\
 \hat{I}^{(2)}_{\Delta}(X^T\rightarrow Y^T)  \triangleq  \sum_{\delta=[0:5:70]  }\hat{I}_{\delta}(\mathcal{X} \rightarrow \mathcal{Y}),\label{sum_estimator}
 \end{align} 
 where $\hat{I}_{\delta}$ is defined according to \eqref{eq1} for any $\delta>0$: 
 \begin{align}
\hat{I}_{\delta}(\mathcal{X} \rightarrow \mathcal{Y})  
\triangleq &  \frac{1}{T}\sum_{t=1}^T I(Y_t;X^{t-\delta}|Y^{t-1}) \label{eqform1}\\
 = &\frac{1}{T}\sum_{t=1}^T \sum_{y_t} \hat{P}(Y_t=y_t\bigl |X^{t-\delta}_{t-\delta-2}=x^{t-\delta}_{t-\delta-2}, Y^{t-1}_{t-2}=y^{t-1}_{t-2})\notag\\
 & \times \log \frac{\hat{P}(Y_t=y_t \bigl| X^{t-\delta}_{t-\delta-2}=x^{t-\delta}_{t-\delta-2}, Y^{t-1}_{t-2}=y^{t-1}_{t-2})}{\hat{P}(Y_t=y_t\bigl|Y^{t-1}_{t-2}=y^{t-1}_{t-2})},\label{eqform2}
\end{align}
and  where $\mathcal{X}$ and $\mathcal{Y}$ denote the (marginal) stationary processes of $X^T$ and $Y^T$. 
Because of the consistency of the initial estimator \eqref{eq1}, it can be checked that \eqref{eqform2} is also consistent provided that
assumptions 1-4 are satisfied.

 
  \subsection{Test on the directed information under fixed stimulation}
 
\begin{figure}[!htbp]\centering
    \includegraphics[width=0.6\columnwidth]{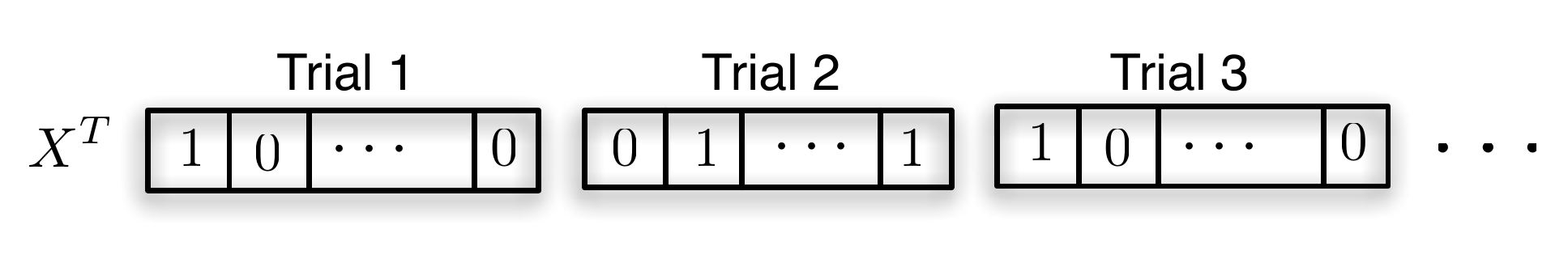}
    \caption[A scheme of the concatenation process]{Trial concatenation (for a given neuron, interval, delay  and frequency pair).}
    \label{fig:spike_trains_concatenate} 
    \end{figure}

We considered correct (also named ``hit") trials recorded for the frequency  pairs $(f1=14,f2=22)$Hz and $(f1=30,f2=22)$Hz. 
  Based on the assumptions of Section \ref{sec:application}, we concatenated all trial segments $x^{T-\delta}$   (respectively $y^T_{\delta+1}$) 
  that were simultaneously  recorded  for every delay $\delta=[0:5:70]$. This concatenation was performed preserving the trial chronology  of each session. 
    For $\delta\geq 0$, this resulted in a  $T'$-length time series, where $T'=(250-\delta) \times \textrm{number of trials}$ bins (See Fig. \ref{fig:spike_trains_concatenate}). 
 
\begin{figure}[!htbp]\centering
  \includegraphics[width=0.6\columnwidth]{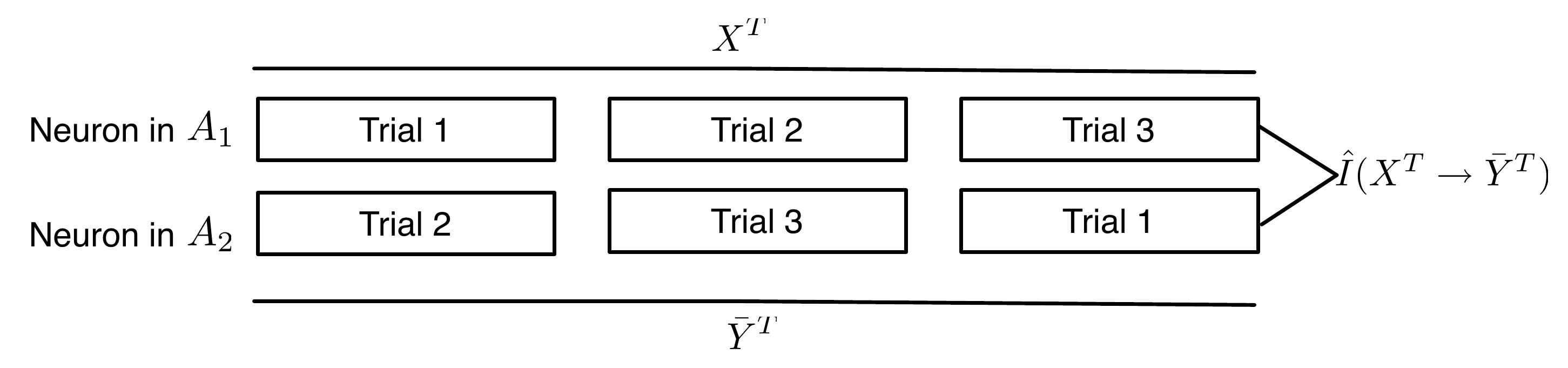}
  \caption[The shuffling procedure]{An example of the permutation procedure between two time series $X^T,\bar{Y} ^{T}$.}
  \label{fig:shuffling} \end{figure}

  To assess the statistical significance of the directed information
associated with each neuron pair and delay we generated surrogate data by permuting $20$ times the concatenation of the second time series  $Y^T$ without replacement (See Fig. \ref{fig:shuffling}). 
 This procedure destroys all simultaneous dependencies but preserves the statistics
 of individual concatenated trials.
Then, we started by testing all single-neuron entropies to determine which neurons were able to express information
about other neurons. Based on this preliminary selection,  
 we tested the (ordered) neuron pairs
 whose endpoint neuron had a significant entropy. 
In more detail, for each delay $\delta=0,5\dotsc, 70$, we thresholded  each original and surrogate data  at  significance level $\alpha=0.05$ by  using a  Monte-Carlo permutation test \cite{Ernst04}, where each value 
 was compared with
 the distribution obtained by adding the original and  the $20$ surrogate estimations.  This gave a number of \emph{thresholded delays} per neuron pair. 
Then, for every neuron pair, we independently tested the estimators  
\eqref{max_estimator} and \eqref{sum_estimator} over all original and surrogate values above the threshold.   
In particular, for the estimator based on the maximization over delays, $\hat{I}^{(1)}_{\Delta}(X^T\rightarrow Y^T)$,  
we used again a Monte-Carlo permutation test \cite{Ernst04}, where this time the original (i.e., non permuted) maximum directed information  
value over thresholded delays  was compared with the tail of
a distribution obtained by aggregating maxima
surrogate values over 
corresponding thresholded delays. 

For the estimator based on the sum of the directed information over delays, $ \hat{I}^{(2)}_{\Delta}(X^T\rightarrow Y^T)$, we summed up the directed information across     
adjacent thresholded delays 
and  used the maximum cluster value as test statistic \cite{Maris07}. Then, we compared the original maximum cluster value with the tail of
a distribution obtained by aggregating maxima
surrogate values over corresponding clusterized delays. 
Significant values of each estimator for either the frequency pair  $(f1=14,f2=22)$Hz or  $(f1=30,f2=22)$Hz defined the responsive paths discussed in the main text.

\begin{figure}[!htbp]\centering
  \includegraphics[width=1\columnwidth]{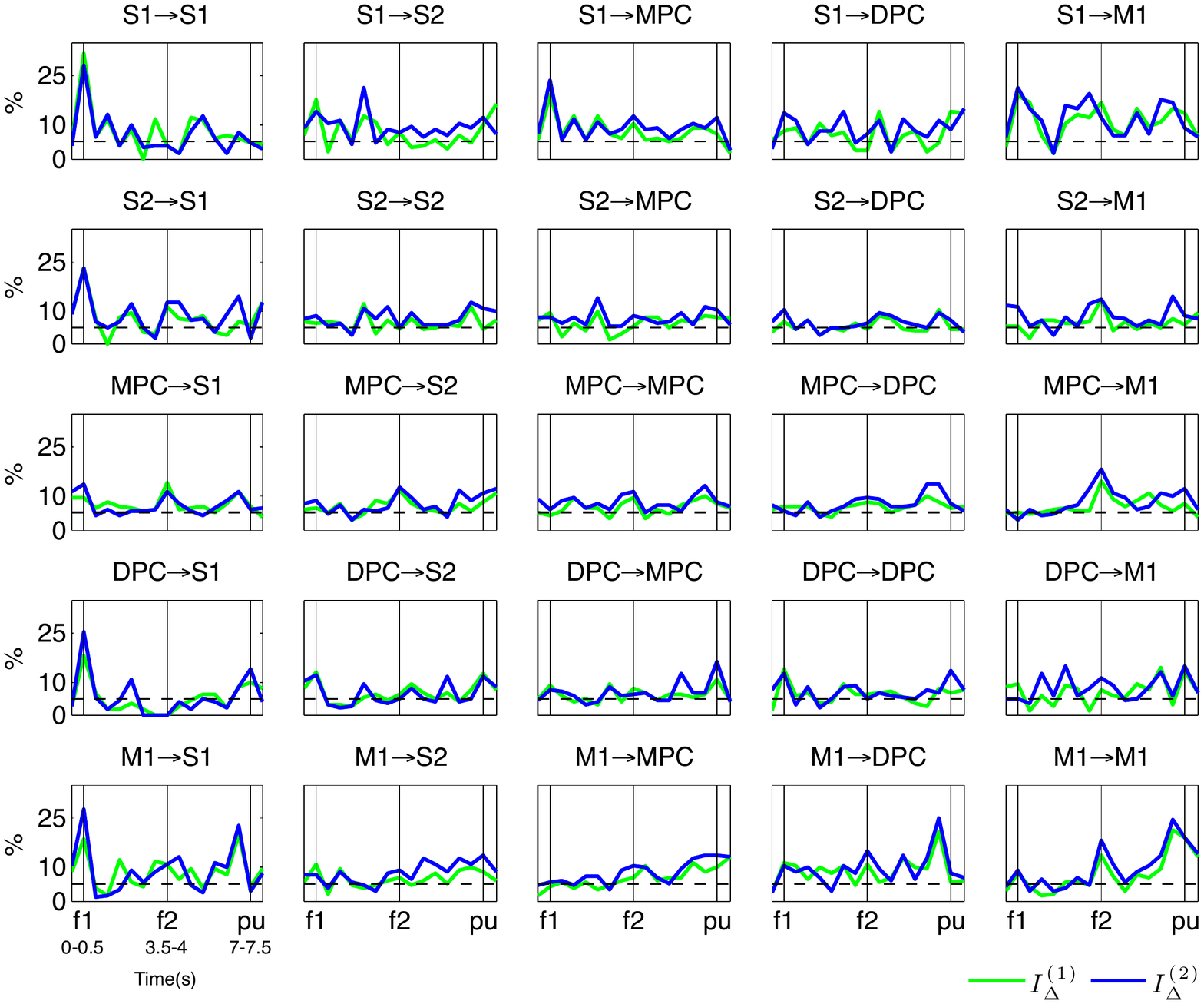}
  \caption{ Comparison of the percentage of modulated paths over responsive paths across all intra- and interarea comparisons between the two proposed directed information 
  estimators in the first monkey. One estimator is based on the maximum directed information over delays (in green) and the other based on the sum of the 
  directed information over delays (in blue). The mean difference is used as a modulation test statistic. 
   Arrows in the title indicate the directionality of the modulated paths.  Vertical bars outline the intervals $f1$, $f2$ and \emph{pu} period.  Horizontal dashed lines indicate the significance level ($\alpha=5\%$).
Data were  obtained in $13$ sessions ($n=13$) from areas
   S1, primary somatosensory cortex; S2, secondary somatosensory cortex; MPC, medial premotor cortex;  DPC, dorsal premotor cortex; M1, primary motor cortex,
    and were plotted for $17$ consecutive  intervals. }
  \label{Figsum} \end{figure}

In order to perform  a specific analysis of interneuronal delays, we chose  $\hat{I}^{(1)}_{\Delta}(X^T\rightarrow Y^T)$ \eqref{max_estimator}
 as our main estimator.  
 Nonetheless, the results using   $\hat{I}^{(2)}_{\Delta}(X^T\rightarrow Y^T)$ \eqref{sum_estimator} were similar as Fig. \ref{Figsum} illustrates.

 \subsection{Test on the modulation of the directed information}

To asses the modulation of the directed information with respect to the frequency sign  $D=f1-f2$, we performed a permutation test for every ordered pair whose directed information had been shown to be significant for either the frequency pair  $(f1=14,f2=22)$Hz or  $(f1=30,f2=22)$Hz with the estimators \eqref{max_estimator}-\eqref{sum_estimator} respectively.
For these pre-selected pairs we  computed directed information 
estimates 
using $5$ trials of each frequency sign. Then, we independently computed  the difference between the median and the mean directed information  
across each set of trials, i.e., $(f1=14,f2=22)$Hz and  $(f1=30,f2=22)$Hz,  as test statistics. For each statistic we 
compared the original value (i.e., non permuted) with the tails of a 
reference 
distribution obtained by permuting   $251$ $\Bigl({10\choose 5} -1\Bigr)$ times the $10$ trials without replacement. 
Significant values were obtained at the two-tailed level $\alpha=0.05$ and defined the modulated paths discussed in the main text. 
The main results of the paper are based on the difference between the means as test statistic, but no relevant differences were found 
using the median. 

   \newpage

\bibliographystyle{IEEEtran}
\bibliography{Bibliography}